\documentclass{article}

\usepackage[final]{neurips_2025}

\usepackage[utf8]{inputenc} 
\usepackage[T1]{fontenc}    
\usepackage{hyperref}       
\usepackage{url}            
\usepackage{booktabs}       
\usepackage{amsfonts}       
\usepackage{nicefrac}       
\usepackage{microtype}      
\usepackage{xcolor}         

\usepackage{graphicx}
\usepackage{subfigure}
\usepackage{algorithm}
\usepackage[noend]{algorithmic}
\usepackage{amsmath}
\usepackage{amssymb}
\usepackage{mathtools}
\usepackage{amsthm}
\usepackage{epsfig}
\usepackage{xspace}
\usepackage{multirow}
\usepackage{listings}
\usepackage{comment}
\usepackage{enumitem}
\usepackage[normalem]{ulem}

\definecolor{lightblue}{rgb}{0.36,0.63,0.90}  
\definecolor{deepblue}{rgb}{0.12,0.31,0.64}

\theoremstyle{plain}

\theoremstyle{definition}

\theoremstyle{remark}

\newenvironment{denseenum}{
\begin{enumerate}[topsep=2pt, partopsep=0pt, leftmargin=1.5em]
  \setlength{\itemsep}{2pt}
  \setlength{\parskip}{0pt}
  \setlength{\parsep}{0pt}
}{\end{enumerate}}

\newenvironment{denseitemize}{
\begin{itemize}[topsep=2.5pt, partopsep=0pt, leftmargin=1.5em]
  \setlength{\itemsep}{2.5pt}
  \setlength{\parskip}{0pt}
  \setlength{\parsep}{0pt}
}{\end{itemize}}

\title{HyGen: Efficient LLM Serving via Elastic Online-Offline Request Co-location}

\author{%
  Ting Sun$^{*1}$, Penghan Wang\thanks{Equal contribution.} \hspace{.01cm} $^2$, Fan Lai$^1$ \\
  $^1$ Siebel School of Computing and Data Science, University of Illinois Urbana-Champaign \\
  $^2$ Department of Computer Science, Purdue University \\
  \texttt{suntcrick@gmail.com}, 
  \texttt{wang6199@purdue.edu},
  \texttt{fanlai@illinois.edu}
}

\begin{document}

\maketitle

\begin{abstract}
Large language models (LLMs) have facilitated a wide range of applications with distinct service-level objectives (SLOs), from latency-sensitive online tasks like interactive chatbots to throughput-oriented offline workloads like data synthesis. The existing deployment model, which dedicates machines to each workload, simplifies SLO management but often leads to poor resource utilization.

This paper introduces HyGen, an interference-aware LLM serving system that enables efficient co-location of online and offline workloads while preserving SLOs. HyGen incorporates two key innovations: (1) performance control mechanisms, including a latency predictor to estimate batch execution time and an SLO-aware profiler to quantify latency interference, and (2) SLO-aware offline scheduling policies that maximize serving throughput and prevent starvation. Our evaluation on production workloads shows that HyGen achieves up to 3.9-5.8$\times$ throughput gains over online and hybrid serving baselines, while ensuring latency SLOs. The code of HyGen is publicly available at \url{https://github.com/UIUC-MLSys/HyGen}.

\end{abstract}

\section{Introduction}

Large language models (LLMs) have emerged as transformative tools across diverse domains, handling both latency-critical online requests (e.g., chatbot interactions~\cite{andes,chatgpt}) and throughput-oriented offline tasks (e.g., document summarization~\cite{llm-market-report}). Online serving demands low and stable response times, measured by Time to First Token (TTFT) and Time Between Tokens (TBT), while offline tasks prioritize high throughput and resource utilization, often processing large batches with relaxed latency constraints. The disparity in these requirements has led most production deployments to segregate online and offline serving onto separate clusters to avoid interference~\cite{patel2024splitwise,sheng2023flexgen,stojkovic2024dynamollm,zhao2024blendserve}. 

However, real-world LLM workloads demonstrate significant temporal variations in request load. Our analysis of production traces shows that in addition to the popular diurnal request arrival patterns, online request rates can vary by up to 3$\times$ within minutes (Section~\ref{sec:background}). To meet latency requirements under such bursty loads, service providers have to provision GPU resources for peak demand, as minute-level resource scaling is often impractical due to infrastructure complexity and engineering overhead~\cite{fu2024serverlessllm,yu2025lambda,zeng2025medusa}---leading to substantial underutilization during off-peak hours. 

This resource overprovisioning suggests an opportunity to improve resource efficiency via hybrid serving---co-locating online and offline workloads on the same inference engine instance. By opportunistically padding online requests with offline requests during periods of low online load, the system could maintain high GPU utilization while preserving latency guarantees for online requests. However, realizing this opportunity requires addressing several fundamental challenges.

First, LLM services exhibit diverse latency requirements across applications~\cite{andes}. Interactive chatbots require consistent response times with both low initial latency (TTFT) and smooth token generation (TBT), while batch processing tasks prioritize throughput over latency. These requirements often manifest in different statistical metrics---from strict P99 latency bounds to mean performance targets---making it difficult to establish unified resource-sharing policies.

Second, LLM workloads are inherently unpredictable in both their arrival patterns and resource demands. 
Request arrival rates exhibit both diurnal patterns and unpredictable short-term fluctuations.
This variability is further complicated by uncertainty in resource demands---input sequences vary widely in length, and the number of output tokens can hardly be predicted until generation completes.

Third, co-locating online and offline workloads introduces interference. For instance, offline requests may delay time-sensitive online requests; using large batch sizes to improve the throughput of offline requests can increase the latency of online requests. Effectively managing this interference while ensuring latency guarantees demands meticulous orchestration of resource sharing.

This paper presents HyGen, an interference-aware LLM serving system that elastically co-locates online and offline workloads. HyGen introduces several key techniques: (1) a latency predictor that accurately estimates the execution time of different request batches, (2) an interference-aware profiler that quantifies the performance interference of co-location, and (3) an adaptive scheduler that maximizes offline throughput while maintaining strict latency guarantees for online requests. 

Our evaluation on production workloads shows that HyGen improves serving throughput by 3.87-5.84$\times$ over existing advances~\cite{sarathi}, while guaranteeing strict SLO compliance.
To summarize, this paper makes the following contributions:
\begin{denseitemize}
    \item Key insights on the feasibility and benefits of co-locating online and offline LLM workloads, derived from systematic characterization of production traces.

    \item A statistical latency prediction model that accurately captures the relationship between batch composition and execution latency, accounting for quadratic complexity in prefill and linear scaling in decode phases.
    
    \item A novel scheduling system that dynamically co-locates online and offline workloads, formulated as a constrained optimization problem that maximizes throughput with prefix sharing while preserving strict latency SLOs and providing theoretical guarantees for fairness.
    
    \item Experimental evaluation on real-world workloads demonstrating up to 5.84$\times$ throughput improvement over state-of-the-art alternatives.
\end{denseitemize}

\section{Related Work}
\label{sec:related}

\paragraph{LLM Inference Optimization.}
Recent advances in LLM inference through kernel optimization~\cite{flashattention,flashinfer}, compilation frameworks~\cite{ma2025intelligen,wu2025mirage,zheng2023einnet}, and scheduling algorithms~\cite{lin2024infinite,lin2024parrot,sun-etal-2025-disco} have substantially improved LLM serving performance. Among these, a particularly important development is predictive scheduling, which aims to estimate request difficulty~\cite{ding2024hybrid, ong2024routellm} or generation length~\cite{fu2024efficient, hu2024inference, jin2023s, patke2024queue, qiu2024power}. Contrasting these efforts, HyGen uniquely predicts batch execution time, providing precise control over request interference during co-located workloads. 
Additionally, the concept of request prefix sharing has been explored by \cite{gim2024prompt, juravsky2024hydragen, zheng2024sglang}, where shared prefixes across requests optimize resource use. HyGen takes this a step further by applying a prefix sharing maximization strategy to opportunistically schedule offline requests, utilizing residual capacity from a primary, SLO-bound online workload. Moreover, HyGen introduces a fairness-aware extension to the standard prefix-sharing maximization (PSM) method, addressing the issue of starvation often encountered in naive designs.

In parallel, the increasing demand for offline LLM inference has driven two key trends. First, platforms like Huggingface Accelerate~\cite{hugging_face_accelerate}, DeepSpeed ZeRO-Inference~\cite{aminabadi2022deepspeed}, and FlexGen~\cite{sheng2023flexgen} enable efficient inference on commodity hardware through memory offloading across GPUs, CPUs, and disk. Second, cloud providers, including OpenAI's Batch API~\cite{openai2024batchapi}, have launched specialized services optimized for throughput rather than latency, processing millions of requests daily for large-scale data processing tasks.

\paragraph{Workloads Co-location.}
In data center environments, co-locating latency-sensitive applications with batch applications has been explored to improve resource utilization~\cite{chen2023olpart, chen2019parties, patel2020clite, tempo-arixv25}. However, these approaches generally overlook the unique characteristics of LLM inference. 
In the context of LLM inference, several works have investigated the co-location of various model types and tasks. For instance, Punica~\cite{chen2024punica}, S-LORA~\cite{sheng2024slora}, and dLoRA~\cite{wu2024dlora} batch requests from different LoRAs, while MuxServe~\cite{duanmuxserve} multiplexes resources across multiple LLMs. In contrast, HyGen focuses on batching online and offline requests, addressing distinct optimization problems that are not interchangeable.

\section{Background and Motivation} \label{sec:background}

This section first introduces the background of LLM serving deployment (Section \ref{sec:llm_inference}), then illustrates how these characteristics motivate our system designs (Section \ref{sec:motivation}).

\subsection{LLM Serving} 
\label{sec:llm_inference}

Large-scale inference clusters consist of multiple serving instances, with a router intelligently directing incoming requests to the most suitable instances~\cite{srivatsa2024preble, iccache-sosp25}. Each instance, which could be based on architectures like vLLM~\cite{kwon2023efficient} or SGLang~\cite{zheng2024sglang}, typically employs iteration-level scheduling~\cite{yu2022orca} and chunked prefill~\cite{sarathi}. This setup enables decode requests to perform an additional decoding step, while prefill requests are limited to a fixed token budget. These two types of requests are processed together within a single iteration, optimizing resource utilization.

LLM serving deployment can be categorized into online serving and offline serving scenarios. 
Online serving targets real-time user interactions, such as chatbots, code assistants, and interactive applications~\cite{githubGitHubCopilot,chatgpt}. This interactive nature often requires a short Time to First Token (TTFT) as well as a short Time Between Tokens (TBT).
Offline serving prioritizes throughput over latency, measured in queries per second (QPS) or tokens per second (TPS). For example, OpenAI's Batch API processes requests with relaxed latency requirements (up to 24 hours) at significantly lower costs compared to standard APIs~\cite{openai2024batchapi}. Applications include model benchmarking~\cite{hendryckstest2021,dubey2024llama3herdmodels}, document processing~\cite{chen2021spreadsheetcoder,zhang-etal-2023-summit}, data cleaning~\cite{narayan2022foundationmodelswrangledata,zheng2023judgingllmasajudgemtbenchchatbot}, and data synthesis~\cite{huggingfaceCosmopediaCreate}.

\subsection{Motivation and Challenges} 
\label{sec:motivation}

\setlength{\textfloatsep}{25pt}

\begin{figure}[t]
    \centering
        \begin{minipage}[t]{0.43\textwidth}
            \centering
            \includegraphics[width=0.8\textwidth]{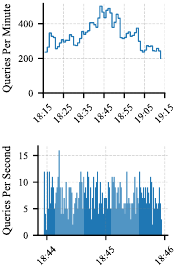}
            \caption{Request rate varies significantly in Microsoft Azure's LLM service over one-hour and two-minute periods.}
            \vspace{-0.2cm}
            \label{fig:azure_code_trace}
        \end{minipage}
        \hfill
        \begin{minipage}[t]{0.55\textwidth}
            \centering
            \includegraphics[width=0.87\textwidth]{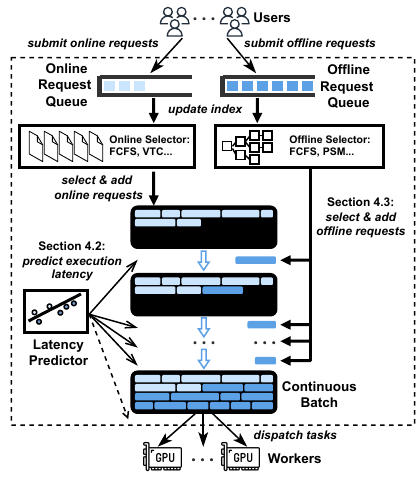}
            \caption{HyGen Overview. Online and offline requests are processed asynchronously, with offline requests opportunistically scheduled to respect latency budgets.}
            \vspace{-0.2cm}
            \label{fig:overview}
        \end{minipage}
\end{figure}

LLM serving systems face a critical resource utilization challenge due to the highly variable nature of their workloads. 
Figure~\ref{fig:azure_code_trace} reports our analysis of Microsoft Azure's LLM service~\cite{patel2024splitwise}, showing that request rates can fluctuate dramatically—varying up to 3× within minutes while following broader diurnal patterns. This variability creates an inherent tension in resource provisioning: serving clusters must be sized to handle peak loads, leading to resource underutilization during off-peak periods.

This load variability suggests an opportunity to improve resource utilization by co-locating online requests with offline requests. For each serving instance in the cluster, during periods of low online traffic, it can opportunistically schedule offline tasks to harvest idle resources. While recent methodologies in simultaneous batching of prefill and decode requests set the premise for dynamic request co-location \cite{sarathi}, doing so at scale introduces several fundamental challenges:

\begin{denseenum}
    \item \textit{Diverse Latency Requirements:} 
    Applications and even requests of an application have distinct latency requirements. For example, paid users require strict latency SLOs while free users accept more relaxed guarantees. 
    How to respect diverse latency requirements in flight?

    \item \textit{Massive Uncertainties:} LLM serving faces temporal uncertainty---online requests arrive in unpredictable bursts with varying urgency levels---as well as resource demand uncertainty due to unpredictable output lengths. 
    How to perform efficient scheduling in the wild?
    
    \item \textit{Request Interference:} 
    Co-locating online and offline workloads introduces performance interference. Large batches of offline requests can cause severe head-of-line blocking, delaying the processing of time-sensitive online requests.
    Worse, batching requests of long inputs with short interactive queries can elongate the latency of all requests in the batch by an order of magnitude~\cite{sarathi}. How to account for interference in co-locating requests?
\end{denseenum}
\section{The HyGen Design}
\label{sec:design}

\subsection{Overview}
\label{sec:design_overview}

HyGen introduces a novel approach to integrate online and offline requests while maintaining strict latency guarantees. As shown in Figure~\ref{fig:overview}, HyGen employs a dual-queue architecture that separates latency-sensitive and throughput-oriented requests. This design accommodates diverse SLOs and variable workloads while remaining compatible with existing scheduling policies within each queue. We note that HyGen functions as an instance-level scheduler, receiving requests from an upstream system-level router (e.g., Preble~\cite{srivatsa2024preble}). As a result, both the request concurrency and scheduling overhead at each instance are inherently bounded.
HyGen's two-phase scheduling operates as follows (see Appendix~\ref{app:two_phase_scheduling} for the asynchronous two-queue workflow and message passing details):

\begin{denseitemize}

 \item The online phase prioritizes latency-sensitive requests, forming an initial batch using established policies such as First-Come-First-Serve (FCFS)~\cite{kwon2023efficient} or fairness request scheduling~\cite{sheng2024fairness}. We introduce a priority-based preemption mechanism that protects online request performance by selectively preempting offline requests. Currently, HyGen preserves execution state for preempted requests, while its architecture supports various preemption mechanisms—including state discarding, preservation, and swapping—as categorized by InferCept~\cite{abhyankar2024infercept}.

\item The offline phase uses our latency predictor to allocate remaining capacity to throughput-oriented requests. This predictor accurately estimates the latency impact of each potential offline request addition, determining either decode request latency costs or maximum chunked prefill lengths that fit within the available latency budget without violating online SLOs.

\end{denseitemize}

Algorithm~\ref{alg:hygen_scheduler} formalizes our scheduling approach, with implementation details and complexity analysis ($O(n)$ where $n$ is the number of requests) provided in Appendices~\ref{app:two_phase_scheduling} and \ref{app:scheduling_complexity}. Appendix~\ref{app:scaling} presents a novel cluster serving paradigm that addresses the longstanding tradeoff between SLO compliance and resource utilization based on HyGen.

\setlength{\textfloatsep}{10pt}
\begin{figure}[t]
\small
\begin{algorithm}[H]
  \caption{HyGen SLO-aware scheduler}
  \label{alg:hygen_scheduler}
  \begin{algorithmic}[1]
  \FUNCTION{SLO\_AWARE\_SCHEDULE}{}
  \STATE \textbf{Input:} running requests $R$, request queue $Q$, 
  \STATE \quad latency budget $t$, chunk size $c$, memory budget $m$
  \STATE \textbf{Output:} batched requests $B$
  \STATE $B \leftarrow \{\}$
  \FOR{$r \in R$.decode}
      \STATE $t_{req} \leftarrow$ PREDICTOR.predict($r$, DECODE) \hspace{.5cm} \textcolor{deepblue}{// predict latency of the decoding request}
      \IF{$t_{req} \leq t$ or PHASE $==$ ONLINE}
          \STATE \textcolor{deepblue}{// schedule request if it is: 1. online, or 2. offline and enough latency budget left}
          \STATE $t \leftarrow t - t_{req}$
          \STATE $B \leftarrow B \cup$ \{($r$, $0$, $t_{req}$)\}
      \ENDIF
  \ENDFOR
  \FOR{$r \in R$.prefill $\cup Q$}
      \STATE \textit{TRY\_SCHEDULE}: \hspace{.5cm} \textcolor{deepblue}{// try to schedule a prefilling or waiting request}
      \STATE \textcolor{deepblue}{// get the max number of tokens allowed under memory and latency budget}
      \STATE $l, t_{req} \leftarrow$ PREDICTOR.get\_max\_tokens($t$, $c$, $m$, $r$) 
      \IF{$l > 0$}
          \STATE \textcolor{deepblue}{// schedule request}
          \STATE $t \leftarrow t - t_{req}$
          \STATE $c \leftarrow c - l$
          \STATE $m \leftarrow m - $ GET\_NUM\_BLOCKS($l$)
          \STATE $B \leftarrow B \cup$ \{($r$, $l$, $t_{req}$)\}
      \ELSE
        \IF {PHASE $==$ ONLINE and $R \neq B$}
          \STATE PERFORM\_PREEMPTION($R$, $m$) \hspace{.5cm} \textcolor{deepblue}{// preempt request with lower priority}
          \STATE \textbf{goto} \textit{TRY\_SCHEDULE} \hspace{.5cm} \textcolor{deepblue}{// try to schedule again}
        \ELSE
          \STATE \textbf{break}
        \ENDIF
      \ENDIF
  \ENDFOR
  \STATE \textbf{return} $B$
  \ENDFUNCTION
  \end{algorithmic}
\end{algorithm}
\end{figure}

\subsection{Performance Control Mechanisms}\label{sec:perf-control}

Co-locating online and offline requests under diverse SLO requirements requires precise control over resource allocation.
The key challenge lies in accurately estimating the latency impact of scheduling decisions to ensure SLO compliance. This section introduces a latency predictor for estimating batch execution latency and an SLO-aware profiler for translating the estimates into scheduling decisions.

\paragraph{Latency Predictor.}
The design of our latency predictor is guided by three key requirements. First, it must provide fast inference to support real-time scheduling decisions. Second, it needs to be robust across varying workload patterns to maintain reliable performance. Third, it should be adaptable to different hardware configurations to accurately capture their unique performance characteristics.

The execution time of an LLM serving batch is primarily determined by two distinct processing stages with different computational patterns. The prefill stage exhibits quadratic complexity due to attention computations, with latency growing quadratically with input sequence length. The total load of this stage depends on both the number of requests and their individual sequence lengths. In contrast, the decode stage shows linear scaling, with computational requirements growing proportionally with the number of tokens.
We can model the batch execution time below to capture these characteristics:
\begin{equation}
    T_{batch} = f(S_p, S_d, S_p^2, S_d^2, N_p, N_d)
\end{equation}
where $S_p$ and $S_d$ represent the total number of prefill and decode tokens in the batch, respectively. The quadratic terms ($S_p^2$ and $S_d^2$) account for non-linear scaling effects, particularly in the prefill phase. $N_p$ and $N_d$ represent the number of prefill and decode requests in the batch, respectively.

We employ linear regression as the prediction model because of its efficiency and effectiveness. Training data for the model is collected by systematically profiling target hardware across diverse batch compositions, varying in the number of requests in different phases, sequence length distributions, and total batch sizes. The linear model enables rapid evaluation of varying batch compositions during scheduling, while its simple feature set ensures stable predictions across varying conditions. Further discussion on the implementation and expandability of the LR predictor can be found in Appendix~\ref{app:predictor_discussion}. 

\paragraph{SLO-aware Profiling.}
The latency predictor provides accurate latency estimates for filling offline requests. Our SLO-aware profiler leverages a latency budget to ensure SLO compliance in scheduling.
The profiler first analyzes the given combination of workload and SLO to establish viable latency budget ranges.
Given that larger batch sizes and longer inputs will increase latency, the profiler test-runs latency budgets within the range to check their compliance with the given SLO and employs binary search to decide an upper limit that meets the overall SLO for online requests. During deployment, this latency budget is used as the batch latency limit in the two-phase scheduling process (Section~\ref{sec:design_overview}) to ensure SLO compliance.
This profiling enables three key capabilities: (1) It determines appropriate latency thresholds that maintain SLO compliance for various workloads and limitations (e.g., power constraints~\cite{patel2024splitwise}). (2) It provides flexible adaptation by adjusting budgets based on changing workload characteristics and performance requirements. (3) It establishes a robust foundation for hybrid scheduling by accounting for both online and offline workload patterns.

\subsection{SLO-aware Offline Scheduling Policies}\label{sec:offline_policies}

After scheduling online requests, our offline scheduling policy repurposes the residual capacity to maximize throughput while ensuring fairness~\cite{sheng2024fairness}. To further optimize the serving throughput for offline requests, HyGen employs an SLO-aware Prefix Sharing Maximization (PSM) strategy. 
Prefix sharing is a widely adopted technique for reusing the KV cache of shared input prefixes between requests~\cite{gim2024prompt,zhao2024blendserve,zheng2024sglang,zheng2024batchllm}. 

\paragraph{Prefix Sharing Maximization Strategy.}\label{sec:prefix_design}  

Our PSM strategy organizes offline requests into a prefix tree following the structure of a Trie tree with each leaf node representing a request, capturing prefix sharing characteristics of all offline requests. The priority of each request is determined by the Depth-First Search (DFS) order of the prefix tree, where requests with the greatest prefix sharing potential are scheduled together. Subsequently, HyGen performs SLO-aware offline scheduling (Section~\ref{sec:design_overview}) using this order to maximize prefix cache reuse, reducing redundant computation and improving throughput. The prefix tree structure also ensures fast insertion and deletion in runtime scheduling. A formalized algorithm is in Appendix~\ref{app:prefix_sharing}. For example, consider a system that can process two offline requests per batch with the following request queue: (\textit{{\color{deepblue}What is} ML}, \textit{{\color{lightblue}How to} code}, \textit{{\color{deepblue}What is} AI}, \textit{{\color{lightblue}How to} debug}). Under traditional FCFS scheduling, requests are processed in arrival order: (\textit{{\color{deepblue}What is} ML}, \textit{{\color{lightblue}How to} code}), (\textit{{\color{deepblue}What is} AI}, \textit{{\color{lightblue}How to} debug}), resulting in no prefix sharing opportunities. In contrast, PSM's prefix-aware scheduling reorders requests as: (\textit{{\color{deepblue}What is} ML}, \textit{{\color{deepblue}What is} AI}), (\textit{{\color{lightblue}How to} code}, \textit{{\color{lightblue}How to} debug}), enabling KV cache reuse through shared prefixes. 

The PSM strategy demonstrates strong extendability. Under certain scenarios, the vanilla PSM strategy may lead to starvation for requests with minimal prefix-sharing potential.
Consider a request queue: (\textit{{\color{deepblue}What is} ML}, \textit{{\color{deepblue}What is} AI}, \textit{{\color{lightblue}How to} code}, \textit{{\color{deepblue}What is} DL}). When new requests arrive with similar prefixes (\textit{{\color{deepblue}What is} LLM}, \textit{{\color{deepblue}What is} DNN}), a naive prefix-sharing policy would continuously prioritize requests sharing the \textit{{\color{deepblue}What is}} prefix, potentially starving the \textit{{\color{lightblue}How to} code} request indefinitely. 
This issue can be mitigated by an extended version of our PSM policy, combining maximum prefix sharing with request freshness by using a utility ratio to ensure a balance between efficiency and fairness. Based on the utility ratio, a new offline request would be selected from the DFS order of the prefix tree or the most stale request from a self-balanced binary search tree sorted by freshness. A detailed algorithm is in Appendix~\ref{app:prefix_sharing_extended}. These enhancements can make the PSM strategy more practical for real-world deployments, retaining its efficiency while improving fairness and adaptability. 
\section{Performance Evaluation}
\label{sec:evaluation}

\begin{figure*}[t]
    \centering
    \includegraphics[width=0.99\textwidth]{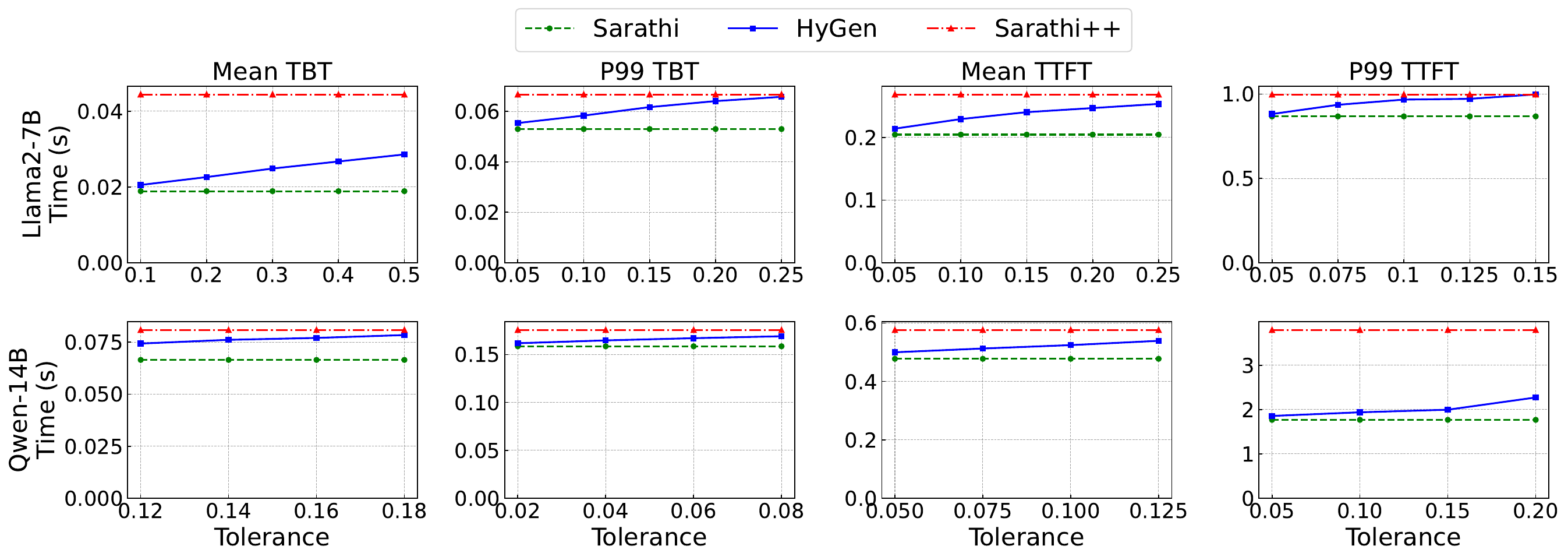}
    \vspace{-.2cm}
    \caption{HyGen respects latency requirements in co-locating requests.}
    \vspace{-.1cm}
    \label{fig:main_latency}
\end{figure*}

\begin{figure*}[t]
    \centering
    \includegraphics[width=0.99\textwidth]{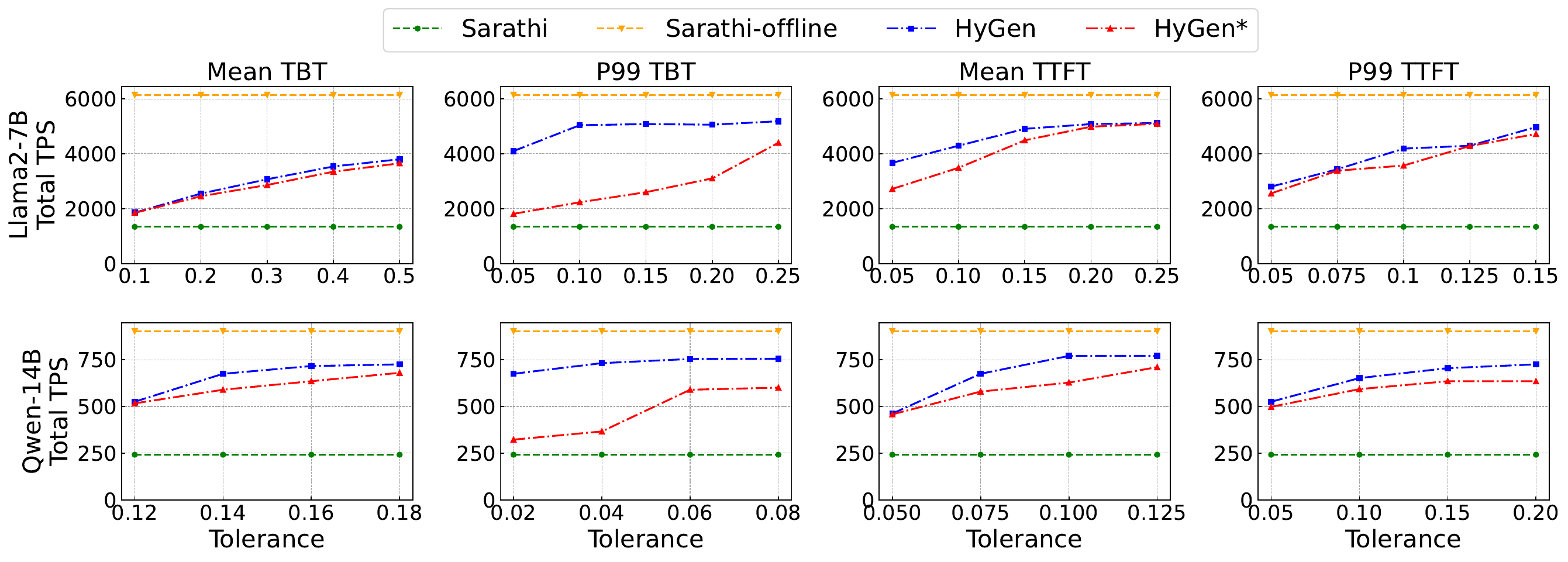}
    \vspace{-.2cm}
    \caption{HyGen improves serving throughput under varying SLOs.}
    \vspace{.1cm}
    \label{fig:main_throughput}
\end{figure*}

\begin{figure}[t]
  \centering
  \begin{minipage}{0.45\columnwidth}
    \begin{center}
    \centerline{\includegraphics[width=0.98\columnwidth]{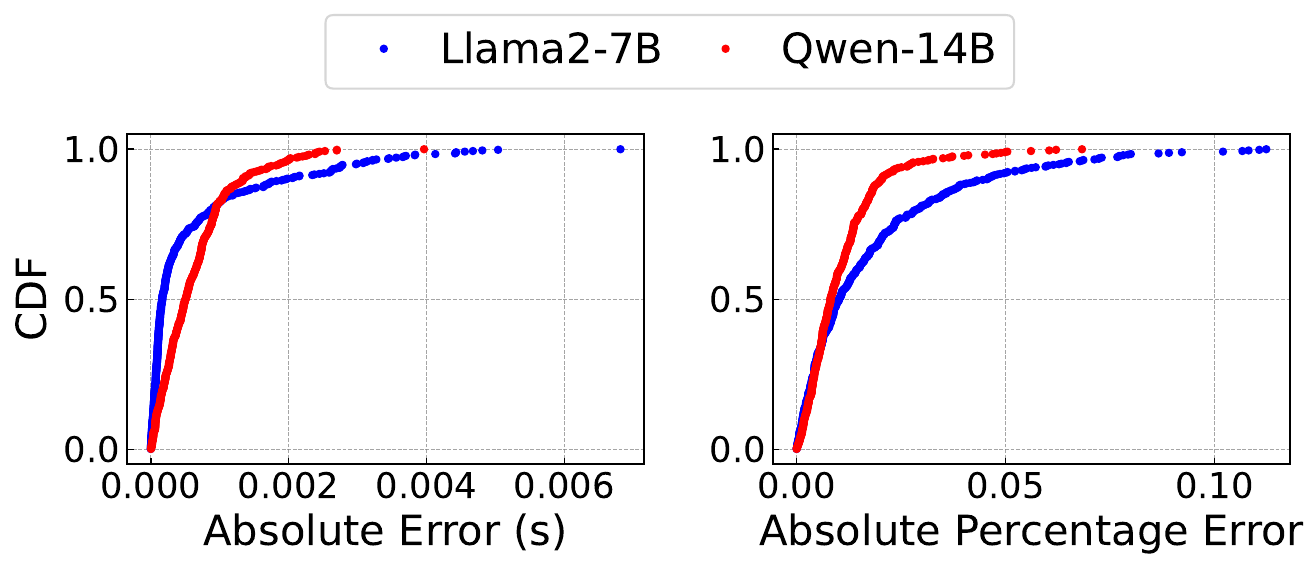}}
    \vspace{.1cm}
    \caption{HyGen latency predictor achieves high accuracy for batch latency prediction.}
    \label{fig:prediction_error}
    \end{center}
  \end{minipage}
  \hfill
  \begin{minipage}{0.26\columnwidth}
    \begin{center}
    \centerline{\includegraphics[width=0.98\columnwidth]{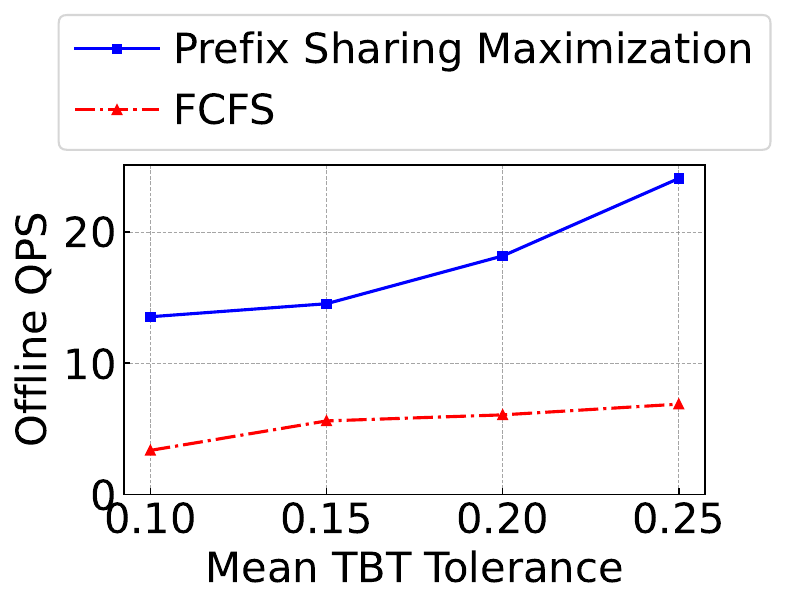}}
    \caption{Prefix Sharing Maximization improves serving throughput.}
    \vspace{-.2cm}
    \label{fig:prefix_sharing}
    \end{center}
    \vspace{-.1cm}
  \end{minipage}
  \hfill
  \begin{minipage}{0.26\columnwidth}
    \begin{center}
    \centerline{\includegraphics[width=\columnwidth]{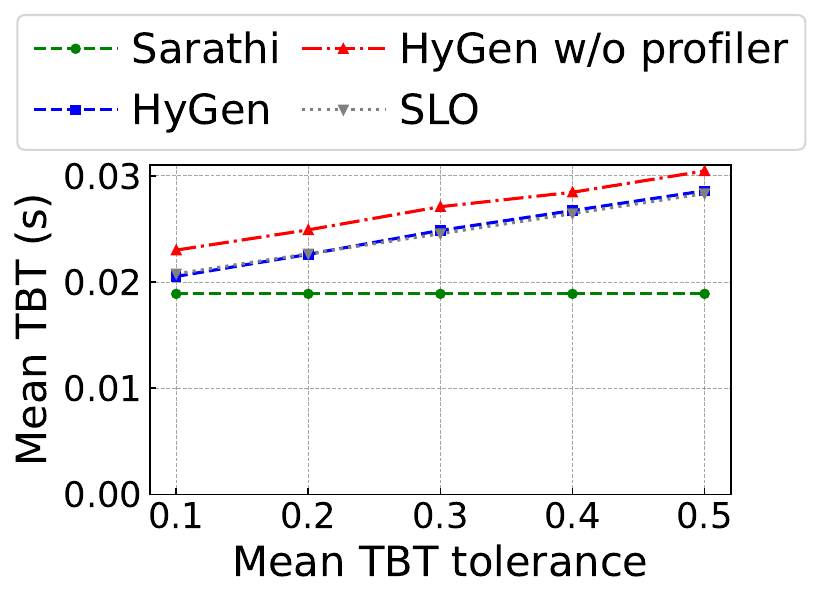}}
    \caption{SLO-aware profiler contributes to HyGen's improvements.}
    \label{fig:breakdown_profiler}
    \vspace{-.2cm}
    \end{center}
  \end{minipage}
\end{figure}

\begin{figure}[t]
    \centering
    \begin{minipage}{0.44\columnwidth}
        \vspace{.2cm}
        \includegraphics[width=0.98\columnwidth]{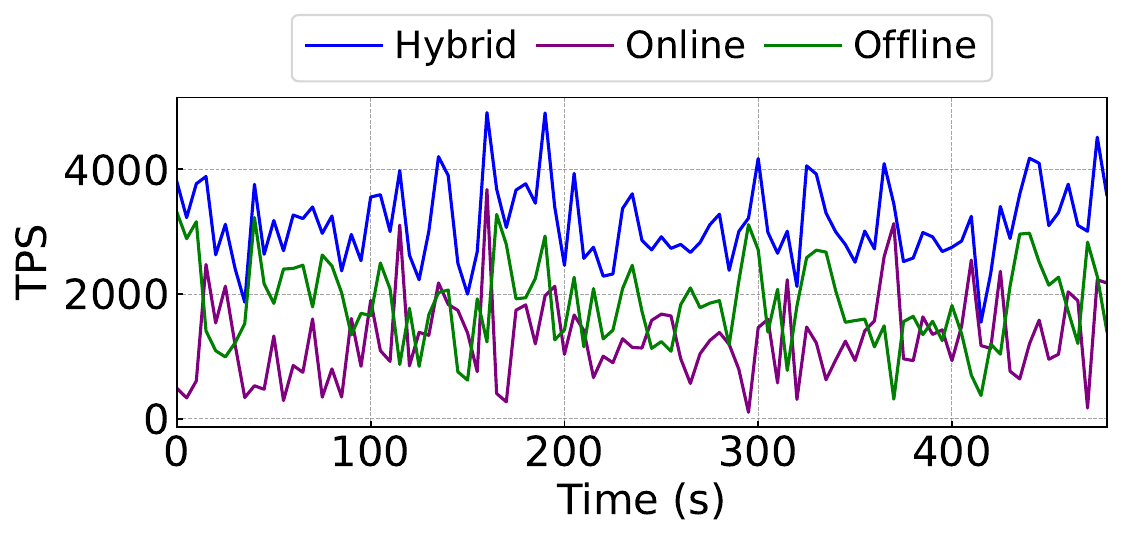}
        \vspace{.2cm}
        \caption{HyGen dynamically controls throughput according to online workload.}
        \label{fig:breakdown_time}
        \vspace{.3cm}
    \end{minipage}
    \hfill
    \begin{minipage}{0.54\columnwidth}
        \vspace{-.4cm}
        \subfigure[SLOs]{\includegraphics[width=0.48\columnwidth]{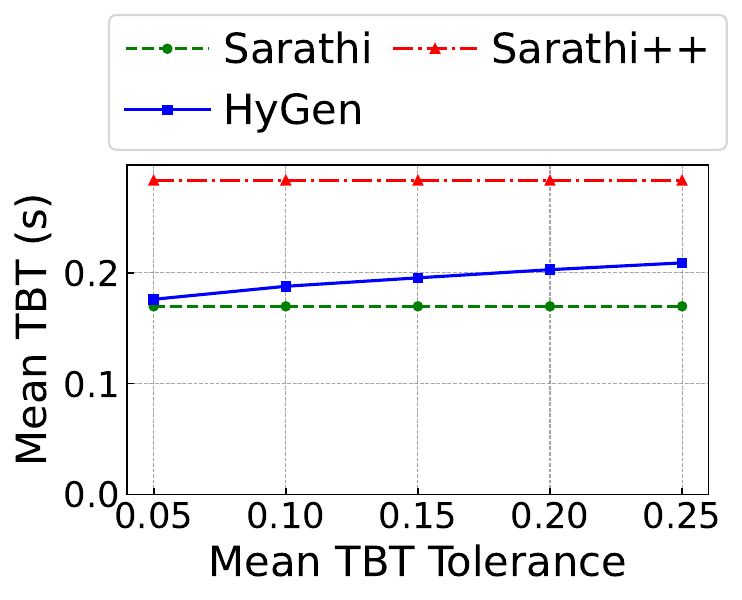}}
        \label{fig:parallel_latency}\hfill
        \subfigure[Throughput]{\includegraphics[width=0.5\columnwidth]{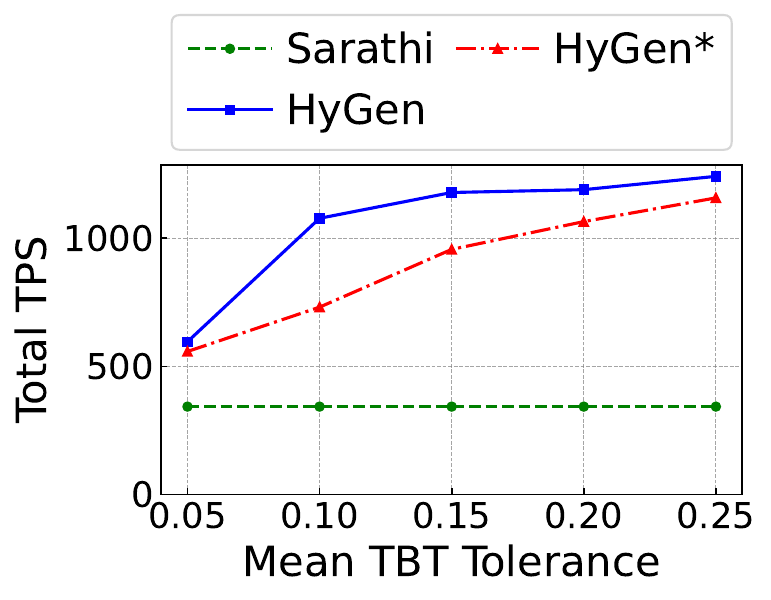}}
        \vspace{-.2cm}
        \caption{HyGen meets SLOs and achieves higher throughput for Yi-34B model using TP=2, PP=2.}
        \vspace{.2cm}
        \label{fig:parallel_exp}
    \end{minipage}
\end{figure}

\begin{figure}[t]
    \centering
    \hspace{-.7cm}
    \includegraphics[width=0.95\textwidth]{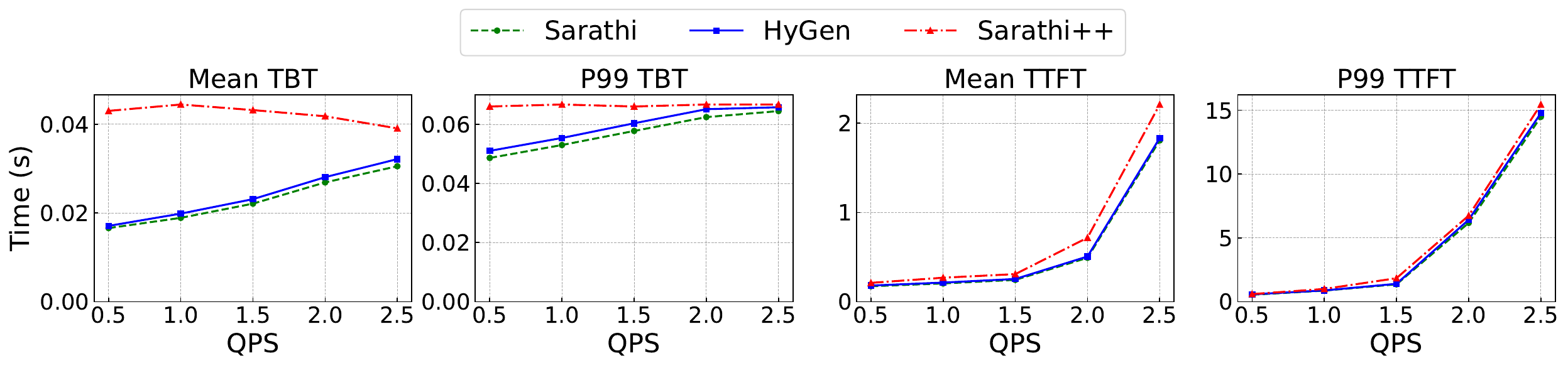}
    \vspace{-.1cm}
    \caption{HyGen meets SLO under various online QPS settings.}
    \vspace{.2cm}
    \label{fig:qps_latency}
\end{figure}

\begin{figure}[t]
    \centering
    \begin{minipage}{0.46\columnwidth}
        \includegraphics[width=\columnwidth]{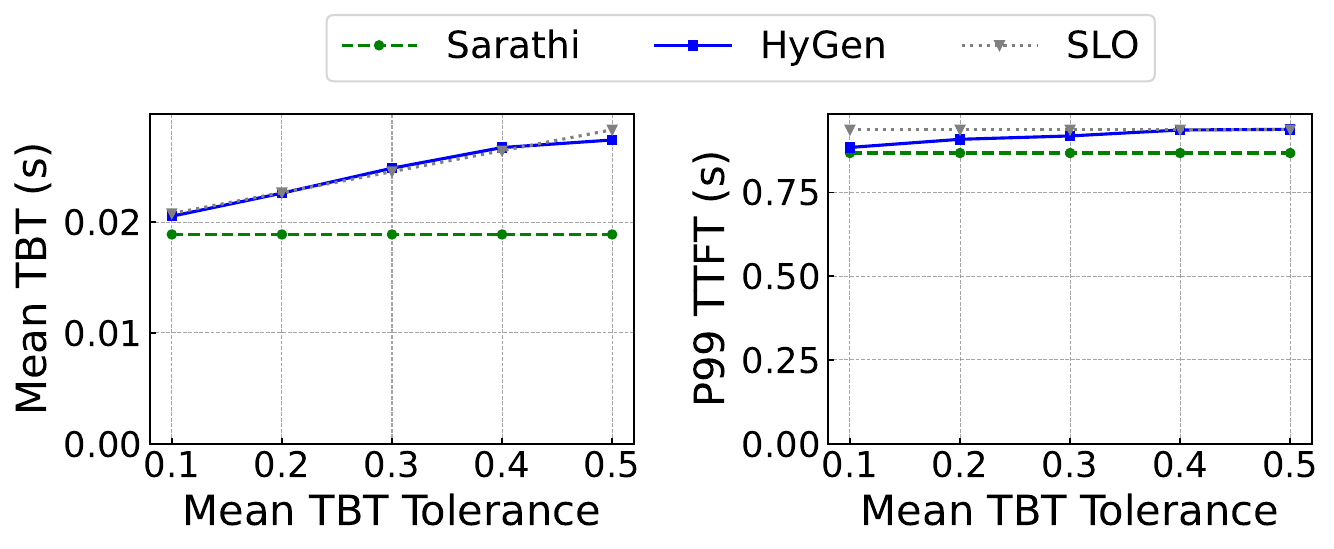}
        \vspace{-.2cm}
        \caption{HyGen is able to meet multiple SLOs simultaneously.
        }
        \vspace{.4cm}
        \label{fig:multi_slo}
    \end{minipage}
    \hfill
    \begin{minipage}{0.52\columnwidth}
        \vspace{-.3cm}
        \subfigure[SLOs]{\includegraphics[width=0.48\columnwidth]{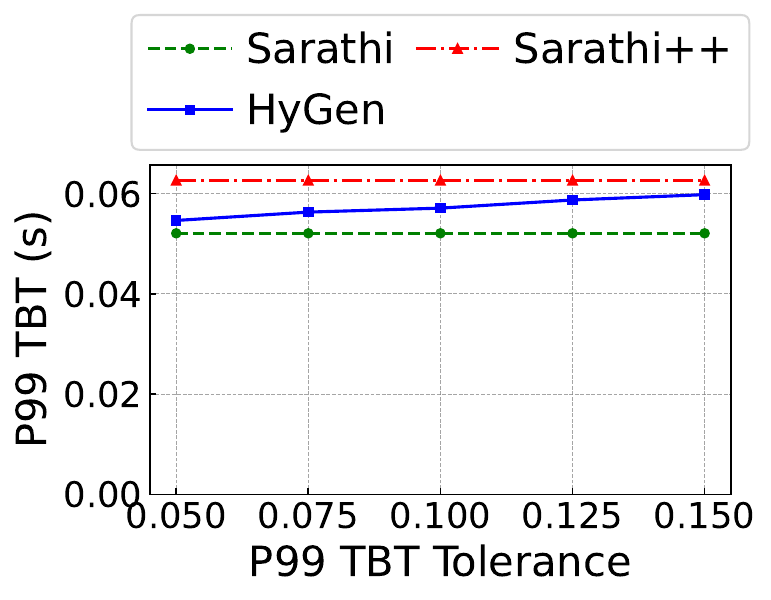}}\hfill
        \subfigure[Throughput]{\includegraphics[width=0.49\columnwidth]{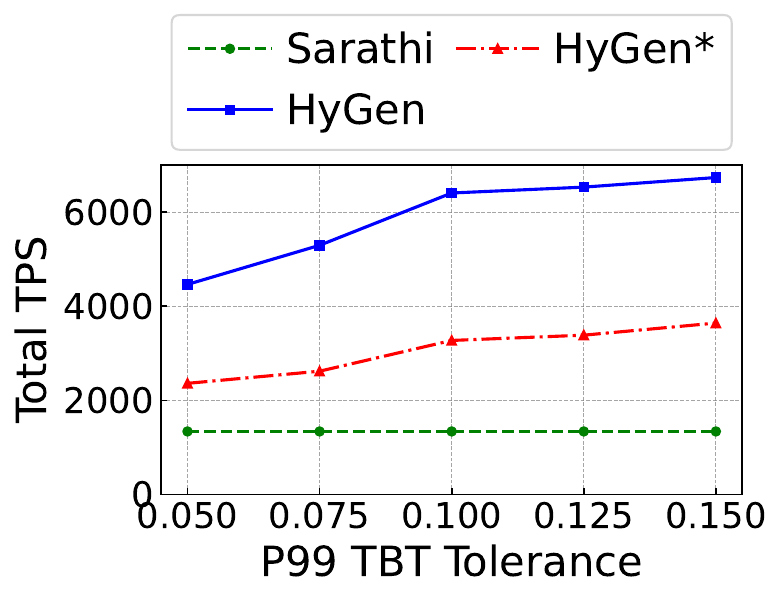}}
        \vspace{-.2cm}
        \caption{HyGen meets SLOs and achieves higher throughput with CNN/DailyMail offline dataset.}
        \vspace{.4cm}
        \label{fig:cnndm}
    \end{minipage}
\end{figure}

\begin{figure}[t]
    \centering
    \begin{minipage}{0.48\columnwidth}
        \subfigure[1 hour]{\includegraphics[width=0.495\columnwidth]{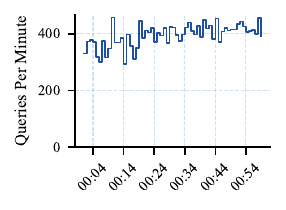}} \hfill
        \subfigure[10 minutes]
        {\includegraphics[width=0.495\columnwidth]{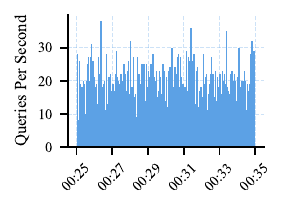}}
        \vspace{-.4cm}
        \caption{Request rate varies in Moonshot Mooncake's LLM service over one-hour and ten-minute periods.}
        \label{fig:mooncake_trace}
    \end{minipage}
    \hfill
    \begin{minipage}{0.48\columnwidth}
        \vspace{-.5cm}
        \subfigure[SLOs]{\includegraphics[width=0.48\columnwidth]{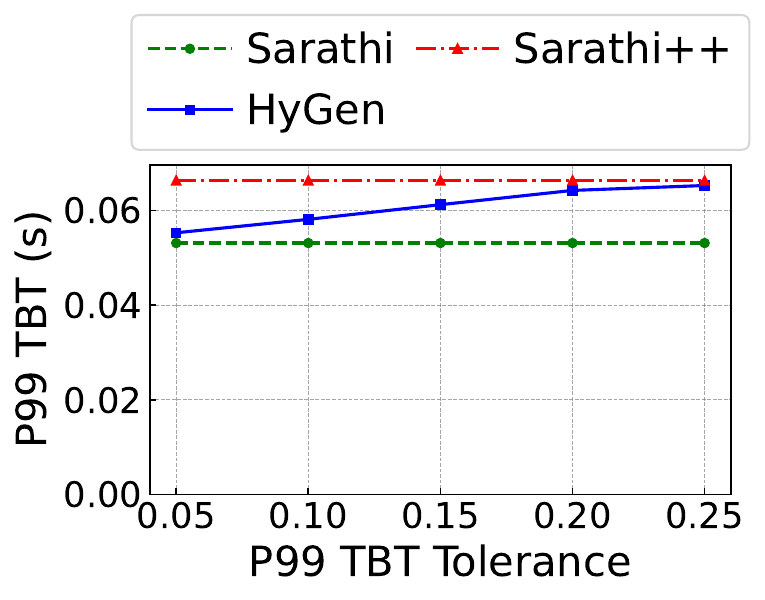}}\hfill
        \subfigure[Throughput]{\includegraphics[width=0.48\columnwidth]{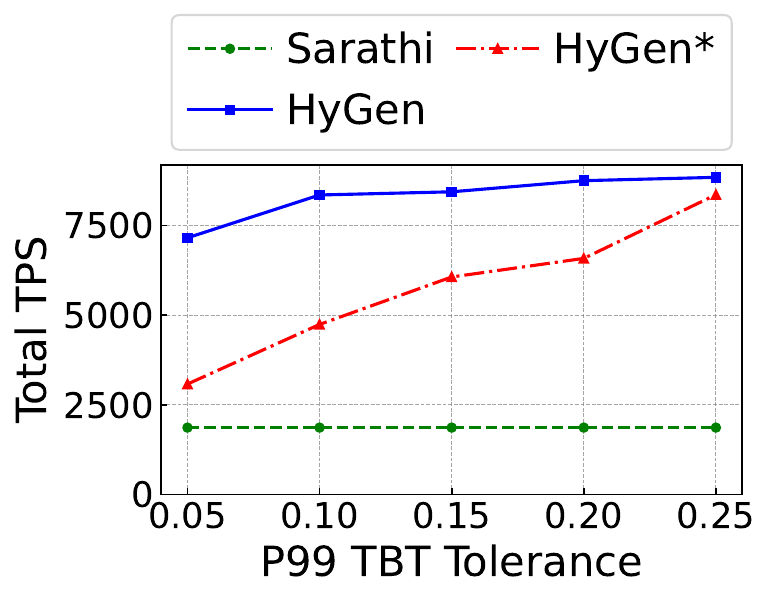}}
        \vspace{-.2cm}
        \caption{HyGen meets SLOs and achieves higher throughput for Mooncake trace.}
        \label{fig:mooncake}
    \end{minipage}
\end{figure}

\begin{figure}[t]
    \centering
    \begin{minipage}{0.48\columnwidth}
        \vspace{-.4cm}
        \subfigure[SLOs]{\includegraphics[width=0.48\columnwidth]{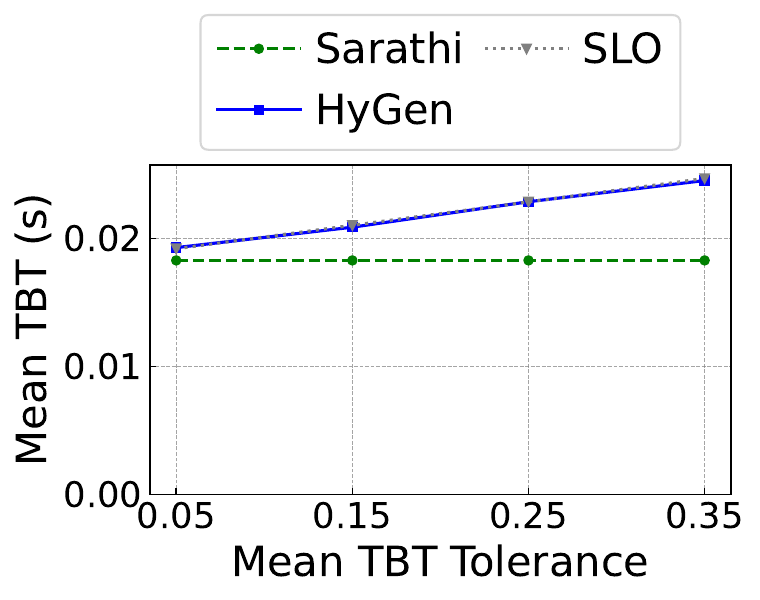}}\hfill
        \subfigure[Throughput]{\includegraphics[width=0.48\columnwidth]{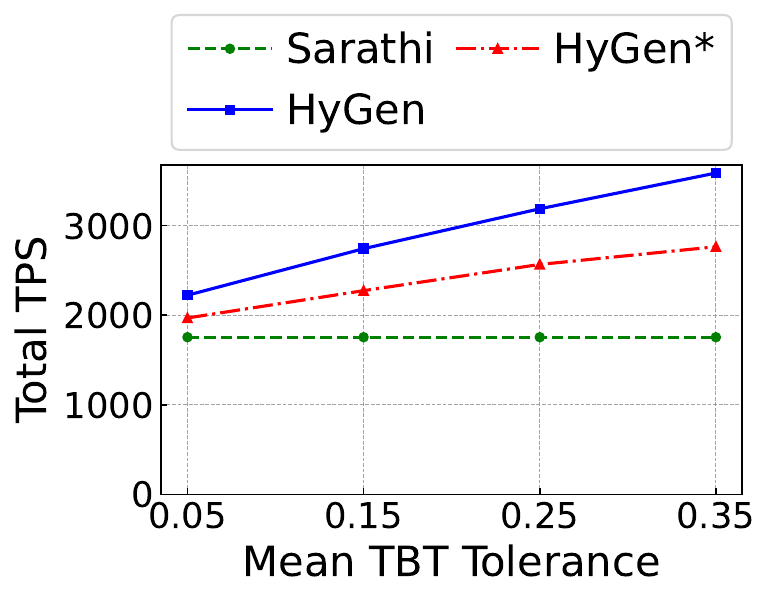}}
        \vspace{-.2cm}
        \caption{HyGen meets SLOs and achieves higher throughput on A5000 GPU and Sheared-LLaMA-2.7B model.}
        \label{fig:a5000}
    \end{minipage}
    \hfill
    \begin{minipage}{0.24\columnwidth}
        \begin{center}
        \vspace{.2cm}
        \centerline{\includegraphics[width=\columnwidth]{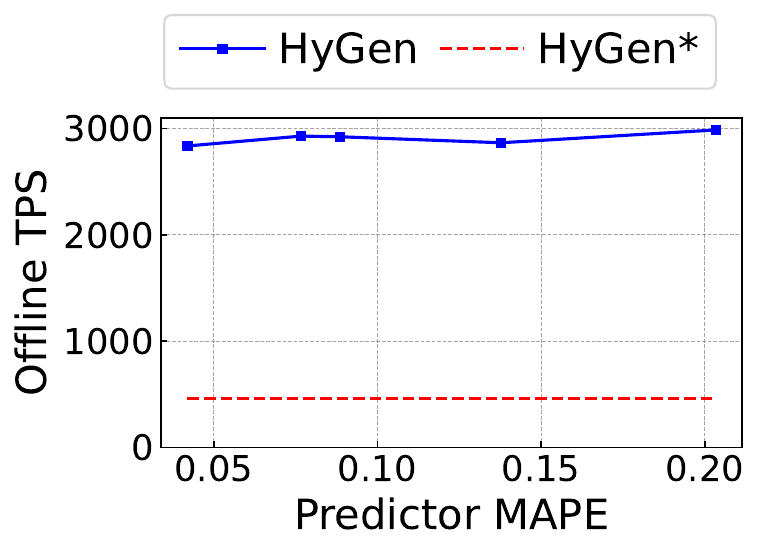}}
        \caption{HyGen is robust to predictor accuracy.}
        \label{fig:ablation_accuracy}
        \end{center}
    \end{minipage}
    \hfill
    \begin{minipage}{0.24\columnwidth}
        \begin{center}
        \vspace{.2cm}
        \centerline{\includegraphics[width=\columnwidth]{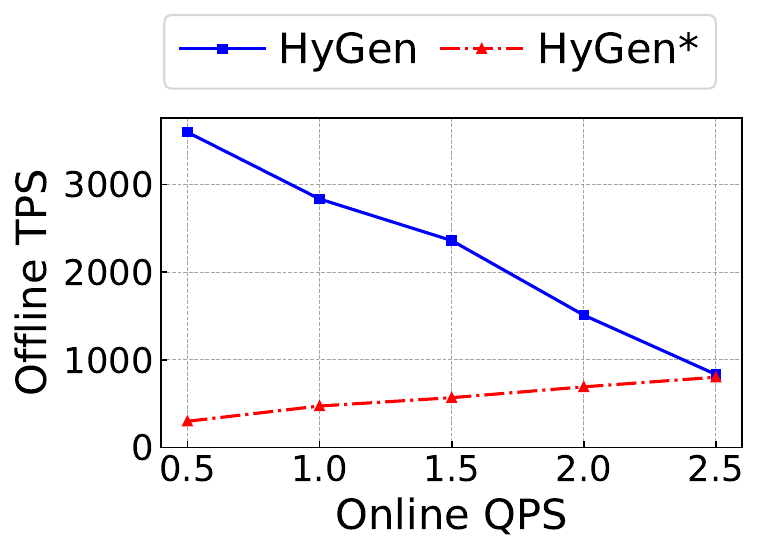}}
        \caption{HyGen dynamically adjusts offline throughput.}
        \label{fig:qps_throughput}
        \end{center}
    \end{minipage}
\end{figure}

\subsection{Evaluation Setup}
\label{sec:evaluation_setup}
\paragraph{Implementation and Testbeds.}
We implement HyGen on top of vLLM \cite{kwon2023efficient,vllm2023github} and Sarathi \cite{sarathi, sarathi2024github}, with 1,300 lines of additional code. 
We evaluate HyGen on three server configurations: one with 4 NVIDIA A100 GPUs (40GB VRAM each), one with 4 NVIDIA A40 GPUs (48GB VRAM each), and one with 1 NVIDIA A5000 GPU (24GB VRAM). All servers have 64 CPU cores, 256GB DDR4 RAM, and a 1.5TB NVMe SSD.

\paragraph{Models and Workloads.} 
For end-to-end evaluation, we use Llama2-7B \cite{touvron2023llama} and Qwen-14B \cite{bai2023qwen} models on A100 and A40 GPUs, respectively. Online workloads are based on the conversation trace from Azure LLM inference trace 2023 \cite{patel2024splitwise}, a one-hour production trace with real-world requests and timestamps. We randomly sampled the trace to achieve the desired QPS that suits our hardware serving capacity. Specifically, within a time duration of $T$ seconds, we would sample $T \times Q$ requests to suit a desired QPS $Q$. For offline workloads, we use arXiv summarization \cite{cohan-etal-2018-discourse}, a dataset for long document summarization.
In our ablation studies, we evaluate HyGen across different model scales ranging from Sheared-LLaMA-2.7B~\cite{xia2023sheared}, Mistral-7B~\cite{jiang2023mistral7b} to Yi-34B~\cite{young2024yi}. The Mooncake trace \cite{mooncake} is further used as the online trace, providing industrial request length distributions and arrival patterns, while CNN/DailyMail~\cite{hermann2015teaching, see-etal-2017-get} and MMLU~\cite{hendrycks2021ethics, hendryckstest2021} are used as offline traces in the ablation studies.
For interference evaluation, we focus on Time to First Token (TTFT) and Time Between Tokens (TBT), including their mean and 99th percentile (P99) values. Throughput is measured in tokens per second (TPS) and queries per second (QPS).

\paragraph{Baselines.}
For pure online inference, we use Sarathi \cite{sarathi} as our baseline. For pure offline serving, we use \textit{Sarathi-offline} to evaluate the maximum offline serving capacity, where an optimal chunk size is profiled for offline workload to maximize throughput. The hyperparameter search of \textit{Sarathi-offline} achieves $\sim$12\% throughput gain compared to the default setup, ensuring optimal baseline performance for fair comparison. We then compare HyGen with two baselines for interference and throughput evaluation, respectively: (1) \textit{Sarathi++}: We implement our online-first scheduling policy on Sarathi to support hybrid serving, including the request management and preemption handling policies introduced in Section~\ref{sec:design_overview}. (2) \textit{HyGen*}: To evaluate the throughput benefit of our HyGen design, we further improve \textit{Sarathi++} to an SLO-aware serving system, \textit{HyGen*}. Besides inheriting the serving policies from \textit{Sarathi++}, \textit{HyGen*} serves offline requests at a specific offline QPS to control overall interference. The offline QPS is profiled using a similar design with the HyGen profiler to guarantee bounded SLO interference.

\subsection{End-to-end Performance}

\textbf{HyGen respects latency requirements in co-locating requests.}
We evaluate HyGen under four SLO metrics (mean TBT, P99 TBT, mean TTFT, and P99 TTFT) with varied interference tolerance ratios. Figure~\ref{fig:main_latency} shows that HyGen controls interference and guarantees to meet specific SLOs across our settings. Compared with \textit{Sarathi++}, an SLO-unaware system that yields the same result for all metrics and tolerance ratios, HyGen shows efficient SLO-aware latency control.

\textbf{HyGen improves serving throughput.}
Figure~\ref{fig:main_throughput} shows the offline throughput of HyGen for various metrics and tolerance ratios. Through efficient request co-location, HyGen improves overall serving throughput by up to 3.87$\times$ compared to pure online serving. Under the same SLO, HyGen consistently achieves higher throughput compared to \textit{HyGen*}, yielding up to $5.84\times$ offline throughput gain. Furthermore, HyGen achieves up to 84.3\% total throughput compared to \textit{Sarathi-offline}, a pure offline serving system whose high throughput benefits from an optimal chunk size profiled for offline requests only. This verifies that our fine-grained latency predictor and SLO-aware profiler designs achieve higher serving efficiency compared to their simplified counterparts.

\subsection{Performance Breakdown}

\paragraph{Accuracy of latency predictor.}
We evaluate the accuracy of our latency predictor on Llama2-7B and Qwen-14B using Azure LLM Inference trace mixed with arXiv summarization dataset. Figure~\ref{fig:prediction_error} shows that our latency predictor achieves a mean absolute percentage error of only 1.78\% and 1.07\%, confirming its high accuracy.

\paragraph{Impact of prefix sharing.}
To test HyGen's compatibility with prefix sharing, we conducted a simulation experiment using Azure LLM Inference as the online trace and MMLU~\cite{hendrycks2021ethics, hendryckstest2021} as the offline dataset on a Llama2-7B model. In our simulation, we deducted the shared prompt prefix length between consecutive offline requests to simulate prefix sharing. Figure~\ref{fig:prefix_sharing} shows that HyGen yields up to $4\times$ offline throughput gain with its prefix sharing maximization scheduling policy.

\paragraph{Impact of SLO-aware profiler.}
To demonstrate the effect of HyGen's SLO-aware profiler, we compared it with a simple strategy that sets the desired mean TBT SLO as the batch latency budget. Figure~\ref{fig:breakdown_profiler} shows the performance gap between individual batch latencies and overall mean TBT, illustrating how the SLO-aware profiler bridges this gap for controlled SLO in hybrid serving.

\paragraph{Breakdown by time.}
Figure~\ref{fig:breakdown_time} shows a temporal throughput breakdown of HyGen. At runtime, HyGen dynamically adjusts offline throughput based on online workload and overall latency budget, batching offline requests more aggressively during online QPS troughs and reducing offline throughput during online bursts, harnessing compute resources in an adaptive manner.

\subsection{Ablation Studies}
\label{sec:evaluation_ablation}

\paragraph{Impact of model parallelisms.}
To evaluate HyGen's effectiveness in distributed inference, we deployed the Yi-34B model \cite{young2024yi} on a server with 4 NVIDIA A40 GPUs using tensor-parallelism (TP) \cite{shoeybi2019megatron} and pipeline-parallelism (PP) \cite{athlur2022varuna,huang2019gpipe,narayanan2019pipedream} with degree 2 for each dimension. Using Azure LLM Inference and arXiv summarization workloads, Figure~\ref{fig:parallel_exp} shows that HyGen maintains its ability to meet SLOs and achieves higher offline throughput (up to $1.89\times$) than the baseline.

\paragraph{Impact of SLO requirements.}
We further evaluate HyGen's ability to meet stringent SLOs under varying online QPS settings on the four aforementioned metrics, each with 5\% interference tolerance. Figure~\ref{fig:qps_latency} shows that HyGen meets stringent SLOs for all metrics. We further demonstrate HyGen's ability to meet multiple SLOs at the same time. By testing HyGen with a fixed P99 TTFT interference ratio (8\%) and mean TBT interference ratios ranging from 10\% to 50\%, Figure~\ref{fig:multi_slo} shows that at a lower mean TBT tolerance, HyGen's performance is bounded by mean TBT SLOs; after reaching the fixed P99 TTFT SLO, mean TBT stops increasing in order to keep P99 TTFT under control.

\paragraph{Impact of models and datasets.}
We further evaluate HyGen's adaptability on two more experiments: The first using Mistral-7B model \cite{jiang2023mistral7b} with Mooncake trace \cite{mooncake}, a trace containing request lengths and timestamps taken from real-world servers, as the online trace, and arXiv summarization as the offline trace; The second experiment uses Llama2-7B model with Azure LLM Inference trace and replaced the offline dataset with CNN/DailyMail summarization dataset \cite{hermann2015teaching, see-etal-2017-get}. Figure~\ref{fig:mooncake_trace} shows the varying request arrival rates of Mooncake trace over one-hour and ten-minute periods, further demonstrating the fluctuating nature of LLM services. Figure~\ref{fig:cnndm} and Figure~\ref{fig:mooncake} show that HyGen achieves superior performance than its counterparts under these settings.

\paragraph{Impact of hardware testbeds.}
To further evaluate HyGen's effectiveness under different hardware configurations, memory limitations and model sizes, we further conducted experiments on A5000 GPU with 24 GB VRAM and Sheared-LLaMA-2.7B model~\cite{xia2023sheared}. Figure~\ref{fig:a5000} shows that HyGen is able to guarantee SLO attainment and achieve higher throughput, with up to 2.18$\times$ offline throughput gain and 1.30$\times$ overall throughput gain compared to the baseline.

\paragraph{Impact of predictor accuracy.}
We tested HyGen's robustness using several pre-trained LR latency predictors with varying prediction accuracy taken from other workloads and tested them on Azure LLM Inference trace and arXiv summarization dataset. Figure~\ref{fig:ablation_accuracy} shows how predictor accuracy (measured in mean absolute percentage error) affects offline throughput under the same P99 TBT SLO. HyGen remains robust across different accuracy settings. Also, our LR-based latency predictor is lightweight for training, with only $\sim$15ms training time for over 80,000 samples on CPUs. HyGen's lightweight latency predictor also only incurs $\sim$18$\mu$s runtime latency per iteration on our experiment CPU, guaranteeing efficient runtime scheduling.

\paragraph{Impact of online arrival rate.} 
Figure~\ref{fig:qps_throughput} shows the effect of online QPS on offline throughput with 5\% P99 TBT tolerance. As online load increases, HyGen adjusts offline throughput based on the system's residual serving capacity while maintaining higher throughputs. 
Understandably, a high online arrival rate limits the headroom for co-location as it approaches system serving capacity. 
\section{Conclusion}
\label{sec:conc}
This paper introduces HyGen, an LLM serving system that enables efficient co-location of online and offline workloads.
We employ control mechanisms to predict and manage interference impacts, and a scheduling policy to opportunistically schedule offline serving. Evaluation on production workloads demonstrates that HyGen improves serving throughput by 3.87-5.84$\times$ while maintaining strict latency SLOs.

\section*{Acknowledgements}
We thank the anonymous reviewers for their constructive and insightful feedback. 
This work was supported in part by grants from Cisco and Google, and by an award from NVIDIA Academic Program. 
It also utilized the Delta system at the National Center for Supercomputing Applications (NCSA) through allocation CIS240236 from the ACCESS program.

\def\UrlBreaks{\do\/\do-}
\bibliography{hygen}

\begin{thebibliography}{10}

\bibitem{abhyankar2024infercept}
Reyna Abhyankar, Zijian He, Vikranth Srivatsa, Hao Zhang, and Yiying Zhang.
\newblock Infercept: Efficient intercept support for augmented large language model inference.
\newblock {\em arXiv preprint arXiv:2402.01869}, 2024.

\bibitem{sarathi}
Amey Agrawal, Nitin Kedia, Ashish Panwar, Jayashree Mohan, Nipun Kwatra, Bhargav~S Gulavani, Alexey Tumanov, and Ramachandran Ramjee.
\newblock Taming throughput-latency tradeoff in llm inference with sarathi-serve.
\newblock In {\em Proceedings of 18th USENIX Symposium on Operating Systems Design and Implementation, 2024, Santa Clara}, 2024.

\bibitem{huggingfaceCosmopediaCreate}
Loubna~Ben Allal, Anton Lozhkov, and Daniel van Strien.
\newblock {C}osmopedia: how to create large-scale synthetic data for pre-training {L}arge {L}anguage {M}odels --- huggingface.co.
\newblock \url{https://huggingface.co/blog/cosmopedia}, 2024.
\newblock [Accessed 25-10-2024].

\bibitem{aminabadi2022deepspeed}
Reza~Yazdani Aminabadi, Samyam Rajbhandari, Ammar~Ahmad Awan, Cheng Li, Du~Li, Elton Zheng, Olatunji Ruwase, Shaden Smith, Minjia Zhang, Jeff Rasley, et~al.
\newblock Deepspeed-inference: enabling efficient inference of transformer models at unprecedented scale.
\newblock In {\em SC22: International Conference for High Performance Computing, Networking, Storage and Analysis}, pages 1--15. IEEE, 2022.

\bibitem{athlur2022varuna}
Sanjith Athlur, Nitika Saran, Muthian Sivathanu, Ramachandran Ramjee, and Nipun Kwatra.
\newblock Varuna: scalable, low-cost training of massive deep learning models.
\newblock In {\em Proceedings of the Seventeenth European Conference on Computer Systems}, pages 472--487, 2022.

\bibitem{bai2023qwen}
Jinze Bai, Shuai Bai, Yunfei Chu, Zeyu Cui, Kai Dang, Xiaodong Deng, Yang Fan, Wenbin Ge, Yu~Han, Fei Huang, et~al.
\newblock Qwen technical report.
\newblock {\em arXiv preprint arXiv:2309.16609}, 2023.

\bibitem{chen2024punica}
Lequn Chen, Zihao Ye, Yongji Wu, Danyang Zhuo, Luis Ceze, and Arvind Krishnamurthy.
\newblock Punica: Multi-tenant lora serving.
\newblock {\em Proceedings of Machine Learning and Systems}, 6:1--13, 2024.

\bibitem{chen2023olpart}
Ruobing Chen, Haosen Shi, Yusen Li, Xiaoguang Liu, and Gang Wang.
\newblock Olpart: Online learning based resource partitioning for colocating multiple latency-critical jobs on commodity computers.
\newblock In {\em Proceedings of the Eighteenth European Conference on Computer Systems}, pages 347--364, 2023.

\bibitem{chen2019parties}
Shuang Chen, Christina Delimitrou, and Jos{\'e}~F Mart{\'\i}nez.
\newblock Parties: Qos-aware resource partitioning for multiple interactive services.
\newblock In {\em Proceedings of the Twenty-Fourth International Conference on Architectural Support for Programming Languages and Operating Systems}, pages 107--120, 2019.

\bibitem{chen2021spreadsheetcoder}
Xinyun Chen, Petros Maniatis, Rishabh Singh, Charles Sutton, Hanjun Dai, Max Lin, and Denny Zhou.
\newblock Spreadsheetcoder: Formula prediction from semi-structured context.
\newblock In {\em International Conference on Machine Learning}, pages 1661--1672. PMLR, 2021.

\bibitem{cohan-etal-2018-discourse}
Arman Cohan, Franck Dernoncourt, Doo~Soon Kim, Trung Bui, Seokhwan Kim, Walter Chang, and Nazli Goharian.
\newblock A discourse-aware attention model for abstractive summarization of long documents.
\newblock In {\em Proceedings of the 2018 Conference of the North {A}merican Chapter of the Association for Computational Linguistics: Human Language Technologies, Volume 2 (Short Papers)}, pages 615--621, New Orleans, Louisiana, June 2018. Association for Computational Linguistics.

\bibitem{flashattention}
Tri Dao, Dan Fu, Stefano Ermon, Atri Rudra, and Christopher R{\'e}.
\newblock Flashattention: Fast and memory-efficient exact attention with io-awareness.
\newblock {\em Advances in Neural Information Processing Systems}, 35:16344--16359, 2022.

\bibitem{ding2024hybrid}
Dujian Ding, Ankur Mallick, Chi Wang, Robert Sim, Subhabrata Mukherjee, Victor Ruhle, Laks~VS Lakshmanan, and Ahmed~Hassan Awadallah.
\newblock Hybrid llm: Cost-efficient and quality-aware query routing.
\newblock {\em arXiv preprint arXiv:2404.14618}, 2024.

\bibitem{duanmuxserve}
Jiangfei Duan, Runyu Lu, Haojie Duanmu, Xiuhong Li, Xingcheng Zhang, Dahua Lin, Ion Stoica, and Hao Zhang.
\newblock Muxserve: Flexible spatial-temporal multiplexing for multiple llm serving.
\newblock In {\em Forty-first International Conference on Machine Learning}, 2024.

\bibitem{fu2024serverlessllm}
Yao Fu, Leyang Xue, Yeqi Huang, Andrei-Octavian Brabete, Dmitrii Ustiugov, Yuvraj Patel, and Luo Mai.
\newblock $\{$ServerlessLLM$\}$:$\{$Low-Latency$\}$ serverless inference for large language models.
\newblock In {\em 18th USENIX Symposium on Operating Systems Design and Implementation (OSDI 24)}, pages 135--153, 2024.

\bibitem{fu2024efficient}
Yichao Fu, Siqi Zhu, Runlong Su, Aurick Qiao, Ion Stoica, and Hao Zhang.
\newblock Efficient llm scheduling by learning to rank.
\newblock {\em arXiv preprint arXiv:2408.15792}, 2024.

\bibitem{gim2024prompt}
In~Gim, Guojun Chen, Seung-seob Lee, Nikhil Sarda, Anurag Khandelwal, and Lin Zhong.
\newblock Prompt cache: Modular attention reuse for low-latency inference.
\newblock {\em Proceedings of Machine Learning and Systems}, 6:325--338, 2024.

\bibitem{llm-market-report}
{Grand View Research}.
\newblock Large language model (llm) market size, share \& trends analysis report by component, by application, by enterprise size, by end-use, by region, and segment forecasts, 2023 - 2030.
\newblock Grand View Research, 2023.

\bibitem{gu2023mamba}
Albert Gu and Tri Dao.
\newblock Mamba: Linear-time sequence modeling with selective state spaces.
\newblock {\em arXiv preprint arXiv:2312.00752}, 2023.

\bibitem{hendrycks2021ethics}
Dan Hendrycks, Collin Burns, Steven Basart, Andrew Critch, Jerry Li, Dawn Song, and Jacob Steinhardt.
\newblock Aligning ai with shared human values.
\newblock {\em Proceedings of the International Conference on Learning Representations (ICLR)}, 2021.

\bibitem{hendryckstest2021}
Dan Hendrycks, Collin Burns, Steven Basart, Andy Zou, Mantas Mazeika, Dawn Song, and Jacob Steinhardt.
\newblock Measuring massive multitask language understanding.
\newblock {\em Proceedings of the International Conference on Learning Representations (ICLR)}, 2021.

\bibitem{hermann2015teaching}
Karl~Moritz Hermann, Tomas Kocisky, Edward Grefenstette, Lasse Espeholt, Will Kay, Mustafa Suleyman, and Phil Blunsom.
\newblock Teaching machines to read and comprehend.
\newblock {\em Advances in neural information processing systems}, 28, 2015.

\bibitem{hu2024inference}
Cunchen Hu, Heyang Huang, Liangliang Xu, Xusheng Chen, Jiang Xu, Shuang Chen, Hao Feng, Chenxi Wang, Sa~Wang, Yungang Bao, et~al.
\newblock Inference without interference: Disaggregate llm inference for mixed downstream workloads.
\newblock {\em arXiv preprint arXiv:2401.11181}, 2024.

\bibitem{huang2019gpipe}
Yanping Huang, Youlong Cheng, Ankur Bapna, Orhan Firat, Dehao Chen, Mia Chen, HyoukJoong Lee, Jiquan Ngiam, Quoc~V Le, Yonghui Wu, et~al.
\newblock Gpipe: Efficient training of giant neural networks using pipeline parallelism.
\newblock {\em Advances in neural information processing systems}, 32, 2019.

\bibitem{hugging_face_accelerate}
{Hugging Face}.
\newblock Hugging face accelerate.
\newblock GitHub repository, 2025.
\newblock Accessed: 2025-01-01.

\bibitem{jiang2023mistral7b}
Albert~Q. Jiang, Alexandre Sablayrolles, Arthur Mensch, Chris Bamford, Devendra~Singh Chaplot, Diego de~las Casas, Florian Bressand, Gianna Lengyel, Guillaume Lample, Lucile Saulnier, Lélio~Renard Lavaud, Marie-Anne Lachaux, Pierre Stock, Teven~Le Scao, Thibaut Lavril, Thomas Wang, Timothée Lacroix, and William~El Sayed.
\newblock Mistral 7b, 2023.

\bibitem{jin2023s}
Yunho Jin, Chun-Feng Wu, David Brooks, and Gu-Yeon Wei.
\newblock S3: Increasing gpu utilization during generative inference for higher throughput.
\newblock {\em Advances in Neural Information Processing Systems}, 36:18015--18027, 2023.

\bibitem{juravsky2024hydragen}
Jordan Juravsky, Bradley Brown, Ryan Ehrlich, Daniel~Y Fu, Christopher R{\'e}, and Azalia Mirhoseini.
\newblock Hydragen: High-throughput llm inference with shared prefixes.
\newblock {\em arXiv preprint arXiv:2402.05099}, 2024.

\bibitem{kwon2023efficient}
Woosuk Kwon, Zhuohan Li, Siyuan Zhuang, Ying Sheng, Lianmin Zheng, Cody~Hao Yu, Joseph Gonzalez, Hao Zhang, and Ion Stoica.
\newblock Efficient memory management for large language model serving with pagedattention.
\newblock In {\em Proceedings of the 29th Symposium on Operating Systems Principles}, pages 611--626, 2023.

\bibitem{lin2024infinite}
Bin Lin, Tao Peng, Chen Zhang, Minmin Sun, Lanbo Li, Hanyu Zhao, Wencong Xiao, Qi~Xu, Xiafei Qiu, Shen Li, et~al.
\newblock Infinite-llm: Efficient llm service for long context with distattention and distributed kvcache.
\newblock {\em arXiv preprint arXiv:2401.02669}, 2024.

\bibitem{lin2024parrot}
Chaofan Lin, Zhenhua Han, Chengruidong Zhang, Yuqing Yang, Fan Yang, Chen Chen, and Lili Qiu.
\newblock Parrot: Efficient serving of $\{$LLM-based$\}$ applications with semantic variable.
\newblock In {\em 18th USENIX Symposium on Operating Systems Design and Implementation (OSDI 24)}, pages 929--945, 2024.

\bibitem{andes}
Jiachen Liu, Zhiyu Wu, Jae-Won Chung, Fan Lai, Myungjin Lee, and Mosharaf Chowdhury.
\newblock Andes: Defining and enhancing quality-of-experience in llm-based text streaming services.
\newblock {\em arXiv preprint arXiv:2404.16283}, 2024.

\bibitem{ma2025intelligen}
Zixuan Ma, Haojie Wang, Jingze Xing, Shuhong Huang, Liyan Zheng, Chen Zhang, Huanqi Cao, Kezhao Huang, Mingshu Zhai, Shizhi Tang, et~al.
\newblock Intelligen: Instruction-level auto-tuning for tensor program with monotonic memory optimization.
\newblock In {\em Proceedings of the 23rd ACM/IEEE International Symposium on Code Generation and Optimization}, pages 107--122, 2025.

\bibitem{dubey2024llama3herdmodels}
Meta-Team.
\newblock The llama 3 herd of models, 2024.

\bibitem{githubGitHubCopilot}
Microsoft.
\newblock {G}it{H}ub {C}opilot · {Y}our {A}{I} pair programmer --- github.com.
\newblock \url{https://github.com/features/copilot}, 2023.
\newblock [Accessed 28-10-2024].

\bibitem{narayan2022foundationmodelswrangledata}
Avanika Narayan, Ines Chami, Laurel Orr, and Christopher R\'{e}.
\newblock Can foundation models wrangle your data?
\newblock {\em Proc. VLDB Endow.}, 16(4):738–746, December 2022.

\bibitem{narayanan2019pipedream}
Deepak Narayanan, Aaron Harlap, Amar Phanishayee, Vivek Seshadri, Nikhil~R Devanur, Gregory~R Ganger, Phillip~B Gibbons, and Matei Zaharia.
\newblock Pipedream: Generalized pipeline parallelism for dnn training.
\newblock In {\em Proceedings of the 27th ACM symposium on operating systems principles}, pages 1--15, 2019.

\bibitem{ong2024routellm}
Isaac Ong, Amjad Almahairi, Vincent Wu, Wei-Lin Chiang, Tianhao Wu, Joseph~E Gonzalez, M~Waleed Kadous, and Ion Stoica.
\newblock Routellm: Learning to route llms from preference data.
\newblock In {\em The Thirteenth International Conference on Learning Representations}, 2024.

\bibitem{chatgpt}
OpenAI.
\newblock Introducing chatgpt.
\newblock \url{https://openai.com/index/chatgpt/}, 2022.
\newblock [Accessed 20-10-2024].

\bibitem{openai2024batchapi}
OpenAI.
\newblock Batch api, 2024.

\bibitem{patel2024splitwise}
Pratyush Patel, Esha Choukse, Chaojie Zhang, Aashaka Shah, {\'I}{\~n}igo Goiri, Saeed Maleki, and Ricardo Bianchini.
\newblock Splitwise: Efficient generative llm inference using phase splitting.
\newblock In {\em 2024 ACM/IEEE 51st Annual International Symposium on Computer Architecture (ISCA)}, pages 118--132. IEEE, 2024.

\bibitem{patel2020clite}
Tirthak Patel and Devesh Tiwari.
\newblock Clite: Efficient and qos-aware co-location of multiple latency-critical jobs for warehouse scale computers.
\newblock In {\em 2020 IEEE International Symposium on High Performance Computer Architecture (HPCA)}, pages 193--206. IEEE, 2020.

\bibitem{patke2024queue}
Archit Patke, Dhemath Reddy, Saurabh Jha, Haoran Qiu, Christian Pinto, Chandra Narayanaswami, Zbigniew Kalbarczyk, and Ravishankar Iyer.
\newblock Queue management for slo-oriented large language model serving.
\newblock In {\em Proceedings of the 2024 ACM Symposium on Cloud Computing}, pages 18--35, 2024.

\bibitem{mooncake}
Ruoyu Qin, Zheming Li, Weiran He, Mingxing Zhang, Yongwei Wu, Weimin Zheng, and Xinran Xu.
\newblock Mooncake: Kimi's kvcache-centric architecture for llm serving.
\newblock {\em arXiv preprint arXiv:2407.00079}, 2024.

\bibitem{qiu2024power}
Haoran Qiu, Weichao Mao, Archit Patke, Shengkun Cui, Saurabh Jha, Chen Wang, Hubertus Franke, Zbigniew Kalbarczyk, Tamer Ba{\c{s}}ar, and Ravishankar~K Iyer.
\newblock Power-aware deep learning model serving with $\{$$\mu$-Serve$\}$.
\newblock In {\em 2024 USENIX Annual Technical Conference (USENIX ATC 24)}, pages 75--93, 2024.

\bibitem{sarathi2024github}
{Sarathi-Serve Project}.
\newblock Sarathi-serve: A low-latency and high-throughput serving engine for llms.
\newblock GitHub repository, 2024.
\newblock Accessed: 2025-01-01.

\bibitem{see-etal-2017-get}
Abigail See, Peter~J Liu, and Christopher~D Manning.
\newblock Get to the point: Summarization with pointer-generator networks.
\newblock In {\em Proceedings of the 55th Annual Meeting of the Association for Computational Linguistics (Volume 1: Long Papers)}, pages 1073--1083, 2017.

\bibitem{shazeer2017outrageously}
Noam Shazeer, Azalia Mirhoseini, Krzysztof Maziarz, Andy Davis, Quoc Le, Geoffrey Hinton, and Jeff Dean.
\newblock Outrageously large neural networks: The sparsely-gated mixture-of-experts layer.
\newblock In {\em International Conference on Learning Representations (ICLR)}, 2017.

\bibitem{sheng2024slora}
Ying Sheng, Shiyi Cao, Dacheng Li, Coleman Hooper, Nicholas Lee, Shuo Yang, Christopher Chou, Banghua Zhu, Lianmin Zheng, Kurt Keutzer, et~al.
\newblock Slora: Scalable serving of thousands of lora adapters.
\newblock {\em Proceedings of Machine Learning and Systems}, 6:296--311, 2024.

\bibitem{sheng2024fairness}
Ying Sheng, Shiyi Cao, Dacheng Li, Banghua Zhu, Zhuohan Li, Danyang Zhuo, Joseph~E Gonzalez, and Ion Stoica.
\newblock Fairness in serving large language models.
\newblock In {\em 18th USENIX Symposium on Operating Systems Design and Implementation (OSDI 24)}, pages 965--988, 2024.

\bibitem{sheng2023flexgen}
Ying Sheng, Lianmin Zheng, Binhang Yuan, Zhuohan Li, Max Ryabinin, Beidi Chen, Percy Liang, Christopher R{\'e}, Ion Stoica, and Ce~Zhang.
\newblock Flexgen: High-throughput generative inference of large language models with a single gpu.
\newblock In {\em International Conference on Machine Learning}, pages 31094--31116. PMLR, 2023.

\bibitem{shoeybi2019megatron}
Mohammad Shoeybi, Mostofa Patwary, Raul Puri, Patrick LeGresley, Jared Casper, and Bryan Catanzaro.
\newblock Megatron-lm: Training multi-billion parameter language models using model parallelism.
\newblock {\em arXiv preprint arXiv:1909.08053}, 2019.

\bibitem{srivatsa2024preble}
Vikranth Srivatsa, Zijian He, Reyna Abhyankar, Dongming Li, and Yiying Zhang.
\newblock Preble: Efficient distributed prompt scheduling for llm serving.
\newblock {\em arXiv preprint arXiv:2407.00023}, 2024.

\bibitem{stojkovic2024dynamollm}
Jovan Stojkovic, Chaojie Zhang, {\'I}{\~n}igo Goiri, Josep Torrellas, and Esha Choukse.
\newblock Dynamollm: Designing llm inference clusters for performance and energy efficiency.
\newblock {\em arXiv preprint arXiv:2408.00741}, 2024.

\bibitem{sun-etal-2025-disco}
Ting Sun, Penghan Wang, and Fan Lai.
\newblock Disco: Device-server collaborative llm-based text streaming services.
\newblock {\em arXiv preprint arXiv:2502.11417}, 2025.

\bibitem{touvron2023llama}
Hugo Touvron, Louis Martin, Kevin Stone, Peter Albert, Amjad Almahairi, Yasmine Babaei, Nikolay Bashlykov, Soumya Batra, Prajjwal Bhargava, Shruti Bhosale, et~al.
\newblock Llama 2: Open foundation and fine-tuned chat models.
\newblock {\em arXiv preprint arXiv:2307.09288}, 2023.

\bibitem{vllm2023github}
{vLLM Project}.
\newblock vllm: Easy, fast, and cheap llm serving with pagedattention.
\newblock GitHub repository, 2023.
\newblock Accessed: 2025-01-01.

\bibitem{wu2024dlora}
Bingyang Wu, Ruidong Zhu, Zili Zhang, Peng Sun, Xuanzhe Liu, and Xin Jin.
\newblock $\{$dLoRA$\}$: Dynamically orchestrating requests and adapters for $\{$LoRA$\}$$\{$LLM$\}$ serving.
\newblock In {\em 18th USENIX Symposium on Operating Systems Design and Implementation (OSDI 24)}, pages 911--927, 2024.

\bibitem{wu2025mirage}
Mengdi Wu, Xinhao Cheng, Shengyu Liu, Chunan Shi, Jianan Ji, Man~Kit Ao, Praveen Velliengiri, Xupeng Miao, Oded Padon, and Zhihao Jia.
\newblock Mirage: A $\{$Multi-Level$\}$ superoptimizer for tensor programs.
\newblock In {\em 19th USENIX Symposium on Operating Systems Design and Implementation (OSDI 25)}, pages 21--38, 2025.

\bibitem{xia2023sheared}
Mengzhou Xia, Tianyu Gao, Zhiyuan Zeng, and Danqi Chen.
\newblock Sheared llama: Accelerating language model pre-training via structured pruning.
\newblock {\em arXiv preprint arXiv:2310.06694}, 2023.

\bibitem{flashinfer}
Zihao Ye, Lequn Chen, Ruihang Lai, Wuwei Lin, Yineng Zhang, Stephanie Wang, Tianqi Chen, Baris Kasikci, Vinod Grover, Arvind Krishnamurthy, et~al.
\newblock Flashinfer: Efficient and customizable attention engine for llm inference serving.
\newblock {\em arXiv preprint arXiv:2501.01005}, 2025.

\bibitem{young2024yi}
Alex Young, Bei Chen, Chao Li, Chengen Huang, Ge~Zhang, Guanwei Zhang, Heng Li, Jiangcheng Zhu, Jianqun Chen, Jing Chang, et~al.
\newblock Yi: Open foundation models by 01. ai.
\newblock {\em arXiv preprint arXiv:2403.04652}, 2024.

\bibitem{yu2022orca}
Gyeong-In Yu, Joo~Seong Jeong, Geon-Woo Kim, Soojeong Kim, and Byung-Gon Chun.
\newblock Orca: A distributed serving system for $\{$Transformer-Based$\}$ generative models.
\newblock In {\em 16th USENIX Symposium on Operating Systems Design and Implementation (OSDI 22)}, pages 521--538, 2022.

\bibitem{yu2025lambda}
Minchen Yu, Rui Yang, Chaobo Jia, Zhaoyuan Su, Sheng Yao, Tingfeng Lan, Yuchen Yang, Yue Cheng, Wei Wang, Ao~Wang, et~al.
\newblock $\{$$\backslash$lambda$\}$ scale: Enabling fast scaling for serverless large language model inference.
\newblock {\em arXiv preprint arXiv:2502.09922}, 2025.

\bibitem{iccache-sosp25}
Yifan Yu, Yu~Gan, Nikhil Sarda, Lillian Tsai, Jiaming Shen, Yanqi Zhou, Arvind Krishnamurthy, Fan Lai, Hank Levy, and David Culler.
\newblock Ic-cache: Efficient large language model serving via in-context caching.
\newblock In {\em SOSP}. ACM, 2025.

\bibitem{zeng2025medusa}
Shaoxun Zeng, Minhui Xie, Shiwei Gao, Youmin Chen, and Youyou Lu.
\newblock Medusa: Accelerating serverless llm inference with materialization.
\newblock In {\em Proceedings of the 30th ACM International Conference on Architectural Support for Programming Languages and Operating Systems, Volume 1}, pages 653--668, 2025.

\bibitem{zhang-etal-2023-summit}
Haopeng Zhang, Xiao Liu, and Jiawei Zhang.
\newblock {S}umm{I}t: Iterative text summarization via {C}hat{GPT}.
\newblock In {\em Findings of the Association for Computational Linguistics: EMNLP 2023}, pages 10644--10657, Singapore, December 2023. Association for Computational Linguistics.

\bibitem{tempo-arixv25}
Wei Zhang, Zhiyu Wu, Yi~Mu, Banruo Liu, Myungjin Lee, and Fan Lai.
\newblock Tempo: Application-aware llm serving with mixed slo requirements.
\newblock {\em arXiv preprint arXiv:2504.20068}, 2025.

\bibitem{zhao2024blendserve}
Yilong Zhao, Shuo Yang, Kan Zhu, Lianmin Zheng, Baris Kasikci, Yang Zhou, Jiarong Xing, and Ion Stoica.
\newblock Blendserve: Optimizing offline inference for auto-regressive large models with resource-aware batching.
\newblock {\em arXiv preprint arXiv:2411.16102}, 2024.

\bibitem{zheng2023judgingllmasajudgemtbenchchatbot}
Lianmin Zheng, Wei-Lin Chiang, Ying Sheng, Siyuan Zhuang, Zhanghao Wu, Yonghao Zhuang, Zi~Lin, Zhuohan Li, Dacheng Li, Eric~P. Xing, Hao Zhang, Joseph~E. Gonzalez, and Ion Stoica.
\newblock Judging llm-as-a-judge with mt-bench and chatbot arena, 2023.

\bibitem{zheng2024sglang}
Lianmin Zheng, Liangsheng Yin, Zhiqiang Xie, Chuyue Sun, Jeff Huang, Cody~Hao Yu, Shiyi Cao, Christos Kozyrakis, Ion Stoica, Joseph~E Gonzalez, et~al.
\newblock Sglang: Efficient execution of structured language model programs.
\newblock {\em arXiv preprint arXiv:2312.07104}, 2024.

\bibitem{zheng2023einnet}
Liyan Zheng, Haojie Wang, Jidong Zhai, Muyan Hu, Zixuan Ma, Tuowei Wang, Shuhong Huang, Xupeng Miao, Shizhi Tang, Kezhao Huang, et~al.
\newblock $\{$EINNET$\}$: Optimizing tensor programs with $\{$Derivation-Based$\}$ transformations.
\newblock In {\em 17th USENIX Symposium on Operating Systems Design and Implementation (OSDI 23)}, pages 739--755, 2023.

\bibitem{zheng2024batchllm}
Zhen Zheng, Xin Ji, Taosong Fang, Fanghao Zhou, Chuanjie Liu, and Gang Peng.
\newblock Batchllm: Optimizing large batched llm inference with global prefix sharing and throughput-oriented token batching.
\newblock {\em arXiv preprint arXiv:2412.03594}, 2024.

\end{thebibliography}
\bibliographystyle{plain}

\newpage

\section*{Limitations}
While HyGen demonstrates substantial improvements in throughput and latency compliance through co-locating online and offline LLM workloads, several limitations remain. First, our approach assumes stable performance predictions from the latency predictor, which may degrade under highly dynamic or adversarial inputs. Second, HyGen focuses on a single model co-location scenario; extending support to heterogeneous models or multi-tenant environments could introduce additional interference patterns. Lastly, our evaluation is limited to specific production workloads—generalizing to other LLM architectures or serving frameworks may require further adaptation and tuning.

\section*{Broader Impact}
This paper proposes HyGen, a LLM serving system for efficient co-location of online and offline requests. Through efficient co-location and SLO control mechanisms, HyGen improves resource utilization and system serving throughput. We believe that the deployment of HyGen will help any kind of LLM service providers by improving serving throughput and providing LLM service with a wider range of options (online/offline). Since our paper provides an efficient serving system for LLM applications without modification to the structure or the outputs of the model, there are no possible negative societal impact that needs to be mentioned in our paper as far as we are concerned.

\section*{Code and Dataset Licenses}

\textbf{Codebase.} HyGen's implementation is based on vLLM\cite{vllm2023github} and Sarathi-Serve\cite{sarathi2024github}, both using Apache-2.0 License.

\textbf{Datasets.} We list the license of used datasets as follows:

arXiv summarization dataset\cite{cohan-etal-2018-discourse}: Apache-2.0 License;

Azure LLM Inference trace\cite{touvron2023llama}: CC-BY-4.0;

MMLU dataset~\cite{hendrycks2021ethics, hendryckstest2021}: MIT License.

\newpage

\appendix
\section*{Appendix}

\section{Algorithms}
\subsection{HyGen Two-phase Scheduling Algorithm}
\label{app:two_phase_scheduling}

This section gives a detailed and formalized demonstration of the two-phase scheduling algorithm introduced in Section~\ref{sec:design_overview}. In each scheduling step, the workflow invokes the SLO-aware scheduling process in Algorithm~\ref{alg:hygen_scheduler} twice (line 13 and line 18) to form a hybrid batch with online and offline request co-location while respecting latency and memory limits. To reduce the scheduling overhead of HyGen, we employ a message queue for asynchronous communication between the main process and the offline scheduler. After each scheduling step, the main process sends the metadata of batched requests to the message queue. The offline scheduler first calculates the expected status of each request based on scheduling decisions from the previous batch, and then runs a scheduling simulation to generate offline request scheduling decisions using our latency predictor and the profiled latency budget. The offline scheduling decisions are then sent back to the main process using the message queue and used for the next scheduling step. To support pipeline parallelization, a scheduling history archive of $K$ steps is kept by the offline scheduler for pipeline parallelization degree $K$, in order to have a holistic view of every request running in each pipeline stage at the time.

\begin{algorithm}[h]
  \caption{HyGen two-phase scheduler}
  \begin{algorithmic}[1]
  \STATE \textbf{global} $Q_{on}$ (online request queue), $Q_{off}$ (offline request queue)
  \STATE \textbf{global} $R_{on}$ (online request running list), $R_{off}$ (offline request running list)
  \STATE \textbf{global} $Q_{send}$, $Q_{recv}$ (message queues)
  \STATE \textbf{global} $L$ (latency budget), $M$ (memory budget), $C$ (chunk size), $M_{off}$ (offline memory)
  \FUNCTION{ASYNC\_SCHEDULER}{}
  \WHILE {True}
      \STATE scheduled requests $S \leftarrow Q_{send}$.get(block=True)
      \STATE UPDATE\_REQUEST\_STATUS($S$, $R_{on}$, $R_{off}$)
      \STATE batched requests $B \leftarrow$ \{\}
      \STATE latency budget $t \leftarrow L$
      \STATE memory budget $m \leftarrow$ GET\_FREE\_MEMORY() + $M_{off}$
      \STATE chunk size $c \leftarrow C$
      \STATE $B \leftarrow B \cup$ SLO\_AWARE\_SCHEDULE($R_{on}$, $Q_{on}$, $t$, $c$, $m$)
      \IF{$m < M_{off}$}
          \STATE PREEMPT\_OFFLINE($R_{off}$, $m$, $M_{off}$)
      \ENDIF
      \STATE $m \leftarrow m - M_{off}$
      \STATE $B \leftarrow B \cup$ SLO\_AWARE\_SCHEDULE($R_{off}$, $Q_{off}$, $t$, $c$, $m$)
      \STATE $Q_{recv}$.send($B$, $t$, $m$)
  \ENDWHILE
  \ENDFUNCTION
  
  \FUNCTION{SCHEDULER}{}
  \STATE \textbf{Output:} batched requests $B$
  \STATE scheduled requests $B$, latency budget $t$, memory budget $m$ $\leftarrow$ $Q_{recv}$.get(block=True)
  \IF{REQUEST\_ARRIVAL}
      \STATE UPDATE\_BUDGET($B$, $t$, $m$)
      \IF{$t < 0$ or $m < 0$}
          \STATE PREEMPT\_UNTIL\_FIT($B$, $t$, $m$)
      \ENDIF
  \ENDIF
  \STATE $Q_{send}$.send($B$)
  \STATE \textbf{return} $B$
  \ENDFUNCTION
  \end{algorithmic}
\end{algorithm}

\newpage

\subsection{SLO-Aware Prefix Sharing Maximization Algorithm}
\label{app:prefix_sharing}

This section details the SLO-aware prefix sharing maximization algorithm design in HyGen. We construct a prefix tree $T_p$ to capture prefix sharing characteristics among offline requests. During scheduling, offline requests are selected in the DFS order of the prefix tree, and deleted once being scheduled. Additionally, running requests keep their original DFS order in future scheduling process, effectively utilizing prefix sharing.

\begin{algorithm}[h]
  \caption{Prefix-sharing-aware offline scheduler}
  \begin{algorithmic}[1]
  \STATE Construct prefix tree $T_p$
  \FUNCTION {PREFIX\_SHARING\_OFFLINE\_SCHEDULE}{}
  \STATE \textbf{Input:} running requests $R$, latency budget $t$, remaining chunk size $c$, memory budget $m$
  \STATE \textbf{Output:} batched requests $B$
  \STATE $B \leftarrow \{\}$
  \FOR{$r \in R$}
      \IF{$r$.state == DECODE}
          \STATE $t_{req} \leftarrow$ PREDICTOR.predict($r$, DECODE)
          \IF{$t > t_{req}$}
              \STATE \textbf{break}
          \ENDIF
          \STATE $t \leftarrow t - t_{req}$
          \STATE $B \leftarrow B \cup$ \{($r$, $0$, $t_{req}$)\}
      \ELSIF{$r$.state == PREFILL}
          \STATE $l, t_{req} \leftarrow$ PREDICTOR.get\_max\_prefill($t$, $c$, $m$, $r$)
          \IF{$l > 0$}
              \STATE $t \leftarrow t - t_{req}$
              \STATE $c \leftarrow c - l$
              \STATE $m \leftarrow m -$ GET\_NUM\_BLOCKS($l$)
              \STATE $B \leftarrow B \cup$ \{($r$, $l$, $t_{req}$)\}
          \ELSE
              \STATE \textbf{break}
          \ENDIF
      \ENDIF
  \ENDFOR
    \WHILE{$T_p$ is not empty}
        \STATE $r \leftarrow T_p$.get\_next\_request()
        \STATE $l, t_{req} \leftarrow$ PREDICTOR.get\_max\_prefill($t$, $c$, $m$, $r$)
        \IF{$l > 0$}
            \STATE $t \leftarrow t - t_{req}$
            \STATE $c \leftarrow c - l$
            \STATE $m \leftarrow m -$ GET\_NUM\_BLOCKS($l$)
            \STATE $B \leftarrow B \cup$ \{($r$, $l$, $t_{req}$)\}
            \STATE $T_p$.remove\_request($r$)
        \ELSE
            \STATE \textbf{break}
        \ENDIF
    \ENDWHILE
  \STATE \textbf{return} $B$
  \ENDFUNCTION
  \end{algorithmic}
\end{algorithm}

\newpage

\subsection{Extended SLO-Aware Prefix Sharing Maximization Algorithm}
\label{app:prefix_sharing_extended}

This section details the extended version of the SLO-aware prefix sharing maximization algorithm, enhanced with fairness-aware scheduling. For offline requests, we construct a prefix tree $T_p$ for prefix sharing, and a self-balancing BST $T_f$ for request freshness. A utility ratio between 0 and 1 is used to balance the chance between these two data structures. During scheduling, the extended prefix sharing maximization algorithm retrieves waiting offline requests from either $T_p$ or $T_f$, based on the utility ratio. The selected request is then deleted from both data structures to ensure synchronization. This solution balances prefix sharing and fairness without disrupting the DFS order of the prefix tree while avoiding possible starvation.

\begin{algorithm}[h]
  \caption{Prefix-sharing-aware offline scheduler}
  \begin{algorithmic}[1]
  \STATE Construct prefix tree $T_p$ and self-balanced BST $T_f$
  \FUNCTION {PREFIX\_SHARING\_OFFLINE\_SCHEDULE}{}
  \STATE \textbf{Input:} running requests $R$, latency budget $t$, remaining chunk size $c$, memory budget $m$, utility value $u$
  \STATE \textbf{Output:} batched requests $B$
  \STATE $B \leftarrow \{\}$
  \FOR{$r \in R$}
      \IF{$r$.state == DECODE}
          \STATE $t_{req} \leftarrow$ PREDICTOR.predict($r$, DECODE)
          \IF{$t > t_{req}$}
              \STATE \textbf{break}
          \ENDIF
          \STATE $t \leftarrow t - t_{req}$, $B \leftarrow B \cup$ \{($r$, $0$, $t_{req}$)\}
      \ELSIF{$r$.state == PREFILL}
          \STATE $l, t_{req} \leftarrow$ PREDICTOR.get\_max\_prefill($t$, $c$, $m$, $r$)
          \IF{$l > 0$}
              \STATE $t \leftarrow t - t_{req}$, $c \leftarrow c - l$, $m \leftarrow m -$ GET\_NUM\_BLOCKS($l$), $B \leftarrow B \cup$ \{($r$, $l$, $t_{req}$)\}
          \ELSE
              \STATE \textbf{break}
          \ENDIF
      \ENDIF
  \ENDFOR
    \WHILE{$T_p$ is not empty}
        \STATE $rand \leftarrow$ RANDOM\_NUMBER(0, 1)
        \IF {$rand < u$}
            \STATE $r \leftarrow T_p$.get\_next\_request()
        \ELSE
            \STATE $r \leftarrow T_f$.get\_next\_request()
        \ENDIF
        \STATE $l, t_{req} \leftarrow$ PREDICTOR.get\_max\_prefill($t$, $c$, $m$, $r$)
        \IF{$l > 0$}
            \STATE $t \leftarrow t - t_{req}$, $c \leftarrow c - l$, $m \leftarrow m -$ GET\_NUM\_BLOCKS($l$), $B \leftarrow B \cup$ \{($r$, $l$, $t_{req}$)\}
            \STATE $T_p$.remove\_request($r$)
            \STATE $T_f$.remove\_request($r$)
        \ELSE
            \STATE \textbf{break}
        \ENDIF
    \ENDWHILE
  \STATE \textbf{return} $B$
  \ENDFUNCTION
  \end{algorithmic}
\end{algorithm}

\newpage

\subsection{Complexity Analysis of the Two-Phase Scheduling Algorithm}
\label{app:scheduling_complexity}

In this section, we present an analysis of the computational complexity of HyGen’s two-phase scheduler. The time complexity of its core components is as follows:

\begin{itemize}
  \item Latency Prediction: $O(1)$ inference using a pre-trained LR model.
  \item PSM: The initial construction of the prefix tree is $O(N×L)$, where $N$ is the number of offline requests and $L$ is the average number of tokens. Each insertion or deletion costs $O(L)$. In implementation, getting the next request in the DFS order only takes $O(1)$, since the DFS order is put in a pre-processed list derived from the prefix tree and can be synced up with the prefix tree asynchronously.
  \item PSM with Fairness: In the fairness-aware PSM algorithm, a self-balancing BST is used for picking the stalest request. Each lookup, insertion, or deletion takes $O(\log n)$. In implementation, the requests can be kept in an FCFS queue, which syncs up with the BST asynchronously to guarantee correctness, so that each lookup still only takes $O(1)$ time.
\end{itemize}

Overall, for the two-phase scheduling using existing policies, the time complexity remains $O(n)$ as asynchronous updates can be performed for advanced policies like fairness-aware PSM.

\section{Further Discussion of the Latency Predictor}
\label{app:predictor_discussion}

Effective co-location in HyGen hinges on a latency predictor that is both highly accurate and computationally lightweight, as it enables real-time scheduling decisions without introducing significant overhead. To this end, we employ a linear regression (LR) model, which provides inference in constant time, $O(1)$, making it ideal for a dynamic serving environment. This section details the model's formulation and its adaptability to the complexities of real-world deployment scenarios.

\paragraph{Model Formulation.}
The execution time of a serving batch, \(T_{\text{batch}}\), is primarily determined by the computational patterns of its prefill and decode stages. We model this relationship as a function of the batch's composition:
\begin{equation}
    T_{\text{batch}} = f(S_p, S_d, S_p^2, N_p, N_d)
    \label{eq:predictor}
\end{equation}
where \(S_p\) and \(S_d\) represent the total number of tokens in the prefill and decode phases, respectively, and \(N_p\) and \(N_d\) are the corresponding request counts. The quadratic term \(S_p^2\) is crucial for capturing the non-linear scaling of the self-attention mechanism, which dominates the computational cost of the prefill stage. In contrast, the decode stage, which processes one token per request at a time, exhibits linear scaling with the batch size (\(N_d\)). The model's coefficients are pre-trained on data gathered by systematically profiling the target hardware and LLM across a diverse set of batch compositions, a lightweight process that ensures applicability to any deployment environment.

\paragraph{Robustness Through System-Level Design.}
While a linear model offers unparalleled efficiency, its accuracy in the face of dynamic system conditions (e.g., GPU temperature variations, resource contention) is a critical consideration. HyGen ensures robustness not through a more complex model, but through a synergistic system design. The SLO-aware profiler (Section~\ref{sec:perf-control}) first establishes a macro-level operational latency budget, which implicitly captures the system's current performance characteristics. The LR predictor then operates at a micro-level, making fine-grained decisions on batch composition \textit{within} this pre-calibrated budget. 

This two-level approach effectively decouples the real-time scheduling decision from low-level hardware variability. As shown in Figure~\ref{fig:ablation_accuracy}, this design allows the system to maintain robust performance and meet SLOs even in scenarios with predictor error rates exceeding 20\%. Furthermore, the feature set is extensible; environmental factors like hardware load can be readily integrated into the model if required for specific use cases.

\paragraph{Adaptability to Modern Model Architectures.}
The feature set defined in Equation~\ref{eq:predictor} is sufficiently general to generalize across various modern LLM architectures without modification. For instance, for Mixture-of-Experts (MoE) models~\citep{shazeer2017outrageously}, where a fixed number of experts are activated per token, the resulting computational cost scales linearly with the total number of tokens processed and is effectively modeled by the \(S_p\) and \(S_d\) features. Similarly, in hybrid architectures that combine linear-complexity components (e.g., Mamba~\citep{gu2023mamba}) with quadratic-complexity attention, our model naturally captures both computational patterns; the linear cost is reflected in the learned coefficient for \(S_p\), while the quadratic cost of the Transformer blocks is captured by the \(S_p^2\) term. This adaptability underscores the predictor's robust design, ensuring its relevance as LLM architectures continue to evolve.

\section{Taming the Throughput-Latency Tradeoff in LLM Serving with HyGen}\label{app:scaling}

Traditional instance scaling solutions address bursty workloads by launching new instances, which can cause tens of seconds to several minutes cold-start delays~\cite{fu2024serverlessllm,yu2025lambda,zeng2025medusa}. To handle these delays, providers often keep standby instances online, leading to wasted resources during off-peak periods.

In contrast, HyGen optimizes resource utilization by running offline workloads on idle resources and reallocating them to online requests in real-time, within a single inference iteration. This eliminates cold-start delays while ensuring high resource utilization.

HyGen complements instance scaling solutions by automating the transition between online and offline workloads, reducing the need for manual intervention. While instance scaling manages large load fluctuations, HyGen ensures efficient resource use during low-traffic periods, optimizing overall system performance in fixed-size clusters.

\end{document}